\documentclass{aa}  
\usepackage{graphicx}
\usepackage[breaklinks, colorlinks, citecolor=blue, linkcolor=blue]{hyperref}
\usepackage{amsmath}
\usepackage{enumitem}
\usepackage{mathtools}
\usepackage{placeins}
\usepackage{subcaption}
\usepackage{caption}
\usepackage{siunitx}
\usepackage{hyperref}
\usepackage{threeparttable}
\usepackage{xcolor}
\usepackage{txfonts}
\usepackage{float}
\usepackage{soul}
\setstcolor{red}

\makeatletter
\DeclareRobustCommand{\ion}[2]{
  \text{#1\,\check@mathfonts\fontsize\sf@size\z@\selectfont #2}
}
\makeatother

\usepackage{xspace}
\newcommand{\blacolor}{red\xspace}
\newcommand{\hicolor}{brown\xspace}
\newcommand{\ovicolor}{green\xspace}
\newcommand{\nlacolor}{blue\xspace}

\newcommand{\kms}{km\,s$^{-1}$\xspace}
\newcommand{\lya}{Ly$\alpha$\xspace}
\newcommand{\linetools}{{\sc linetools}}

\begin{document}

   \title{Pinpointing the cosmic web between massive galaxy clusters} 
    \titlerunning{The incidence of H~{\sc i} and O~{\sc vi} as WHIM tracers in inter-cluster filaments}
   \subtitle{I. The incidence of H~{\sc i} and O~{\sc vi} as WHIM tracers}

   \author{Karen Martínez-Acosta
          \inst{1,2}\corrauth{karen.m91martinez@gmail.com}
          \and
          Nicolas Tejos\inst{1}\corrauth{nicolas.tejos@pucv.cl}
        \and
        Ismael Pessa
        \inst{3}\email{ismael.pessa@gmail.com}
        \and
        J Xavier Prochaska
        \inst{4,5}\email{xavier@ucolick.org}
        \and
        Joseph Burchett
        \inst{6}\email{jnb@nmsu.edu}
        \and
        Britton Smith
        \inst{7}\email{britton.smith@ed.ac.uk}
        \and
        Jessica Werk
        \inst{8}\email{jwerk@uw.edu}
        \and
         Nelson Padilla
         \inst{9}\email{n.d.padilla@gmail.com}
        \and
          Sebastián López
          \inst{10}\email{slopez@das.uchile.cl}
          }

   \institute{Instituto de F\'isica, Pontificia Universidad Cat\'olica de
Valpara\'iso, Casilla 4059, Valpara\'iso, Chile
         \and
             Facultad de Ciencias, Universidad Técnica Federico Santa María, Av. España 1680, Valparaíso, Chile
        \and
            Leibniz-Institut for Astrophysik Potsdam (AIP), An der Sternwarte 16, 14482 Potsdam, Germany
        \and
            Kavli Institute for the Physics and Mathematics of the Universe (WPI), The University of Tokyo, Kashiwa, Chiba 277-8583, Japan
        \and
            Department of Astronomy and Astrophysics, University of California, Santa Cruz, 1156 High Street, Santa Cruz, CA 95064, USA
        \and
            Department of Astronomy, New Mexico State University, Las Cruces, NM 88003, USA
        \and
             Institute for Astronomy, University of Edinburgh, Royal Observatory, Edinburgh, EH9 3HJ, UK
        \and 
            Department of Astronomy, University of Washington, Seattle, WA 98195, USA
        \and
            Instituto de Astronomía Teórica y Experimental, CONICET-Universidad Nacional de Córdoba, Laprida 854, X5000BGR, Córdoba, Argentina
        \and
            Departamento de Astronomía, Universidad de Chile, Camino El Observatorio 1515, Las Condes, Santiago, Chile
             }

   \date{Received Month DD, YYYY; accepted  Month DD, YYYY}

  \abstract 
{Currently at redshift $z<1$, the $\Lambda$CDM paradigm predicts that $\sim40-50\%$ of baryons are located in the warm-hot intergalactic medium (WHIM) within cosmic web filaments. Characterizing the physical state of this dominant baryonic matter and its distribution is crucial for a complete picture of baryon evolution in the low-redshift Universe. However, directly observing the WHIM has been challenging due to its low density and high temperature and ionization state. Absorption line spectroscopy offers a promising route, in particular by analyses of broad \lya absorbers (BLAs; Doppler parameter $b\geq40$ \kms) and \ion{O}{VI} absorbers.} 
{This work focuses on probing the signatures of the WHIM by analyzing the incidence rate of these tracers, in and around putative cosmic web filaments connecting massive galaxy clusters.}
{We employ Far Ultraviolet (FUV) absorption spectroscopy acquired with the HST/COS along ten QSO sightlines at $z<1$ selected based on their proximity (in projected distance at low velocity separation) to cluster-pairs. We focus our analysis on absorption features of WHIM tracers (BLAs and \ion{O}{VI}) and cold gas tracers (narrow \lya absorbers, or NLAs; $b<40$ \kms), which are used as control samples for temperature.}
{We measure an excess in the incidence rates of total \ion{H}{I}, NLAs, BLAs, and \ion{O}{VI} within a rest-frame $\Delta v=\pm1000$ \kms and an impact parameter $\Delta d\leq3$ Mpc relative to the field expectation, finding that the excesses of BLAs and \ion{O}{VI} are approximately twice those of NLAs. Additionally, the covering fraction for BLAs and \ion{O}{VI} appears to be larger (by $\sim1.7$ and $\sim1.8$ times, respectively) than for a random (field) control sample of the environment, whereas the covering fraction for NLAs is consistent with the random expectation.}
{The larger relative excess of WHIM tracers compared to cold gas tracers in the vicinity of independent cluster-pairs indicates the prevalence of warm-hot gas in these environments, representing a potential signature of WHIM detection in inter-cluster filaments.}


   \maketitle
   \nolinenumbers

\section{Introduction}
The warm-hot intergalactic medium (WHIM) is a low-density (n$_H\sim10^{-6}-10^{-4}$ cm$^{-3}$) plasma with temperatures of $T\sim 10^5-10^7$ K and mild overdensities of $\sim$$ 10 - 30$ times the mean density of the Universe (\citealt{Cen1999}, \citealt{Dave2001},
\citealt{Prause2007}, \citealt{Danforth2010}, \citealt{Savage2011}, \citealt{DOdorico2023}, \citealt{Zhao2025}). The WHIM is believed to emerge from intergalactic gas that is shock-heated to warm-hot temperatures as the medium collapses under gravity (\citealt{Cen1999}, \citealt{Dave2001}, \citealt{Richter2006}), by cloud–cloud collisions, active galactic nuclei (AGN) outflows and/or galactic winds (\citealt{Cen2006a}, \citealt{Cen2006b}). 
Analyses of structure formation predict a universe interconnected by a complex network of filamentary structures referred to as the `cosmic web', composed of dark matter and diffuse baryons (\citealt{Bond1996}, \citealt{Dave2001}, \citealt{Tepper2012}, \citealt{Burchett2020}).
The WHIM is expected to reside mainly in the densest filaments of this cosmic web (\citealt{Tejos2016}, \citealt{Galarraga2021}, \citealt{DOdorico2023}, \citealt{Luber2025}).

In the $\Lambda$ Cold Dark Matter ($\Lambda$CDM) framework, galaxy clusters correspond to the nodes of the cosmic web, where several filaments converge. In these models, matter is predicted to flow along these filamentary structures, contributing to the mass assembly of galaxy clusters \citep{Bond1996}. Galaxy cluster outskirts are particularly relevant in this context, as they connect the filamentary large-scale structure of the cosmic web to galaxy clusters (\citealt{Battaglia2015}, \citealt{Kuchner2020}). Galaxy cluster outskirts are complex structures characterized by gas shock processes triggered by substructures, gas and galaxies falling into clusters, and diffuse material collapsing towards filaments \citep{Rost2021}. In the far outskirts of clusters, the intracluster medium (ICM) meets the intergalactic medium (IGM), the gaseous component of the connective tissues of the cosmic web \citep{Holguin2024}. Cosmological dark matter simulations predict high probabilities of having filamentary structures between close ($< 20$ Mpc) and pairs of massive ($> 10^{14}  M_{\odot}$) galaxy clusters (\citealt{Colberg2005}, \citealt{Aragon2010}, \citealt{Gonzalez2010}), so a natural place to search for WHIM signatures and filaments is the volume between pairs of massive galaxy clusters. 

Cosmological simulations have reproduced and analyzed the formation and dynamics of the filamentary structure of the universe, using different models, such as Millennium Simulations (\citealt{Springel2005}), Illustris (\citealt{Vogelsberger2013}, \citealt{Martizzi2019}), EAGLE (\citealt{Schaye2015}), and FLAMINGO (\citealt{Schaye2023}), among many others. From an observational perspective, the filamentary structure of the cosmic web has been a key feature in understanding cosmic evolution (\citealt{deLapparent1986}). In the case of the WHIM, it has been studied using different multi-wavelength methods. X-ray emission and absorption provide information on the spatial distribution of baryons (\citealt{Eckert2015}, \citealt{Nicastro2016}, \citealt{Reiprich2021}, \citealt{Tanimura2022}, \citealt{Zhang2024}, \citealt{Zhao2025}). Observational techniques such as the thermal Sunyaev$-$Zel’dovich (tSZ) effect (\citealt{Bonjean2018}, \citealt{deGraaff2019}, \citealt{Tanimura2020}) have been employed to detect signatures of the WHIM in the filaments of a nearby galaxy cluster. These observational studies frequently report statistical significance at the $3-4$ $\sigma$ level  \citep{Erciyes2023}. Notwithstanding this degree of significance, in the findings of \cite{Bonjean2018}, the limited optical and infrared data prevented the detection of cosmic filaments, a result of particular relevance because most of the WHIM is expected to be located in these filaments (\citealt{Dave2001}, \citealt{Eckert2015}). 
Furthermore, diffuse radio emission has been detected between pairs of clusters (\citealt{Govoni2019}, \citealt{Pignataro2024}), consistent with overdensities corresponding to filaments as well as weak lensing detections between Abell cluster-pairs \citep{Shinde2025}. Given the diffuse nature of the WHIM and its low neutral hydrogen fraction, the photon emission at far ultraviolet (FUV) wavelengths is limited (aside from stacking photons from many different regions); thus, absorption-line measurements currently represent the most suitable method for studying the WHIM at low redshifts (\citealt{Richter2006}, \citealt{Savage2011}, \citealt{Tejos2016}). 

Due to its high temperature and low density, directly detecting the WHIM is challenging. Nevertheless, characterizing this medium is essential to comprehend the local large-scale structure, galaxy feedback, and evolution (\citealt{Lehner2007}, \citealt{Prause2007}, \citealt{Danforth2010}, \citealt{Richter2010}, \citealt{Tejos2016}). For instance, a recent work by \cite{Macquart2020} used the dispersion measure of a sample of localized fast radio bursts and measured the baryon content of the Universe to be $\Omega_{b}= 0.051_{-0.025}^{+0.021} h_{70}^{-1}$. This result implies that the baryons formerly referred to as `missing' (\citealt{Cen1999}, \citealt{Bregman2007}, \citealt{Shull2012}) are likely to reside in the IGM, but it does not reveal their spatial distribution or physical state. 

Quasar far ultraviolet (FUV-QSO) absorption spectroscopy is an observational technique that relies on bright background continuum sources, such as QSOs, to identify and characterize intervening absorption features along the sightline. It offers several advantages over attempts to detect gas using emission spectroscopy (\citealt{Richter2008}, \citealt{Ahoranta2020}). These include that absorption spectroscopy is more sensitive in a diffuse medium because of its linear dependence on the density of the absorbing gas\footnote{As opposed to emissivity, which depends on the square of the density.}, that absorption lines incorporate kinematic characteristics of the gas, and that the ionization state of the gas can be analyzed using multiple ions that trace the same kinematic component (\citealt{Richter2008}, \citealt{Tepper2012}, \citealt{Tejos2016}). Given its physical properties, the WHIM phase is most effectively detected with this technique (\citealt{Shull2003}, \citealt{Danforth2010}, \citealt{Tilton2012}). 

Transitions of \ion{H}{I} and \ion{O}{VI} have been employed to search for evidence of the WHIM in filamentary structures (\citealt{Richter2006}, \citealt{Tejos2016}, \citealt{Ahoranta2020}). The FUV range enables access to the Ly$\alpha$ transition, the strongest and most common for probing the IGM. Having direct coverage of \ion{H}{I}, independent of the presence of metals, is of great value for detecting the WHIM because this medium may remain metal-poor (\citealt{Richter2006}, \citealt{Danforth2010}, \citealt{Tejos2016}). 
Neutral hydrogen generally traces cool and photoionized gas, but it may also trace collisionally ionized gas in the WHIM through thermally broadened lines (\citealt{Tepper2012}, \citealt{Tejos2016}). Among the high-ionization state ions, a key absorption line to trace the WHIM is \ion{O}{VI} because it is a relatively abundant element \citep{Richter2006}. However, the physical conditions of the gas giving rise to \ion{O}{VI} absorption remain uncertain, and no general consensus has been reached \citep{Tepper2011}. \cite{Danforth2008} found that \ion{O}{VI} can reliably trace collisionally ionized gas at temperatures of $T \sim 10^5- 10^6$ K, whereas \cite{Thom2008a} and \cite{Tripp2008} argued that \ion{O}{VI} absorption predominantly arises from photoionized gas at temperatures below $10^5$ K. Nevertheless, these studies also found that more than half of their \ion{O}{VI} absorbers could trace both lower-metallicity collisionally ionized gas at $T > 10^5$ K and cold photoionized gas. Under collisional ionization equilibrium (CIE), \ion{O}{VI} mainly traces gas at $T \sim 3 \times 10^5$ K, making it a sensitive tracer of the lower-temperature regime of the WHIM. At higher temperatures of the WHIM, oxygen is predominantly ionized into \ion{O}{VII} and \ion{O}{VIII}. The corresponding absorption transitions lie in the X-ray regime, where high-quality spectroscopic observations remain relatively limited  (\citealt{Richter2006b}, \citealt{Tepper2011}). 

Other high-ion transitions used to probe the WHIM include those of carbon, silicon, nitrogen, and neon. However, these are generally less informative, either because their accessible transitions preferentially trace gas at lower temperatures or because the abundances of these elements in the IGM are too low to yield robust detections  (\citealt{Lehner2007}, \citealt{Richter2008}, \citealt{Richter2010}, \citealt{Tepper2012}, \citealt{Artale2022}). Notably, \ion{Ne}{VIII} is a promising tracer of the IGM and the circumgalactic medium (CGM) since it is the fifth most abundant element in the Universe. In particular, \ion{Ne}{VIII} reliably traces gas warmed to $ T \sim 4\times 10^5- 2 \times 10^6$ K through collisional ionization (\citealt{Gnat2007}, \citealt{Frank2018}).

The width of the absorption feature, known as the Doppler parameter ($b$), depends on the temperature of the gas. The thermal broadening of a line profile can be used to infer the temperature according to $b_{th}=0.129\sqrt{\frac{T}{A}}$ \kms, where $A$ is the atomic weight of the element, and $T$ is the gas temperature in Kelvin. The non-thermal broadening includes turbulence and unresolved blends. In overdense environments, turbulence is expected to be the dominant source of non-thermal physical broadening, and can be parametrized as proportional to the thermal broadening, such that $b_{\text{non-th}} \approx b_{\text{turb}} \approx \alpha b_{\text{th}}$ (\citealt{Richter2006b}, \citealt{Tepper2012}, \citealt{Pessa2018}, \citealt{Pessa2025}). The Doppler parameter is composed of thermal and non-thermal components according to $b=\sqrt{b_{\text{th}}^2+b_{\text{non-th}}^2}$. \cite{Pessa2025} found that absorbers with $b\geq40$ \kms trace the WHIM at $T>10^5$ K without a critical dependence on the value adopted for the parameter $\alpha$, which accounts for the turbulent contribution\footnote{This is valid for $\alpha<1$. For $\alpha>1$, $T$ may be slightly lower than $10^5$ K; however, these values of $\alpha$ are less likely, given observational samples and simulated BLAs (see \citealt{Pessa2025} for further details).}. Therefore, the identification of broad absorbers is not significantly affected by the assumed level of turbulence. In particular, the Doppler parameter criterion used to define broad Ly$\alpha$ absorbers (BLAs) remains robust under reasonable variations in the turbulent contribution. In this work, we identify as BLA any \ion{H}{I} features with $b \geq 40$ \kms \citep[see, e.g.,][]{Richter2006b, Danforth2010, Danforth2016}.

In this paper, we advance on the pioneering work of \cite{Tejos2016}, which provided a first statistical assessment in the detection of WHIM in inter-cluster filaments of the cosmic web using a single QSO sightline, and established the experimental design and methodology. The search is conducted along QSO sightlines that intersect the transverse line connecting two galaxy clusters at approximately the same redshift. Here, we analyze a survey of ten QSO spectra, representing a significantly larger sample than that employed by \cite{Tejos2016}, based on a single sightline. We estimate the incidence rate of \ion{O}{VI} and \ion{H}{I}, as well as sub-samples of BLAs, and narrow Ly$\alpha$ absorbers (NLAs, defined as \ion{H}{I} features with $b<40$ \kms). We search for WHIM signatures by analyzing the incidence rates of WHIM tracers and comparing them with a control sample of cold gas (NLAs), targeting potential cosmic web filaments that may connect massive galaxy clusters. 

This paper is organized as follows.
Section~\ref{data} describes the data and target selection. In Section~\ref{analysis}, we explain the methodology used to characterize the ten QSO sightlines provided by the cluster-pair sample and to search for absorption line systems. Section~\ref{results} presents the results on the incidence rate of total \ion{H}{I} (narrow and broad absorbers), NLAs, BLAs, and \ion{O}{VI} obtained in our survey. In Section~\ref{Discussion}, we discuss the incidence rate and covering fraction of the absorption features analyzed in our survey. Finally, in Section~\ref{conclusions}, a summary and the conclusions of this study are presented. Here, we use the following cosmological parameters: Hubble constant $H_0 = 67.8$ \kms Mpc$^{-1}$, mass energy density $\Omega_{m} = 0.307$, dark energy density $\Omega_{\Lambda}= 0.691$, spatial curvature $k=0$ (\citealt{Planck2014}), and distances are in co-moving coordinates.

\section{Data}
\label{data}
Our experiment aims to probe filaments that may connect massive galaxy clusters. Accordingly, we analyzed the FUV spectra of background QSOs with putative intervening inter-cluster filaments to construct a catalog of intervening putative filaments associated with QSO sightlines. First, we used a galaxy cluster catalog to identify cluster-pairs to target the densest filaments in the `cosmic web’ (see Section~\ref{2.1}). Then, we cross-matched these cluster-pairs with our UV bright QSO catalog, and the product of this cross-match is our QSO-filament catalog (see Section~\ref{2.2}). The criteria imposed to select the galaxy clusters and QSO samples are the same as those employed in previous works (\citealt{Tejos2016}, \citealt{Pessa2018}, \citealt{Pessa2025}), summarized as follows.

\subsection{Galaxy cluster catalog and cluster-pairs}\label{2.1}

To build our galaxy cluster catalog, we work with the red-sequence Matched-filter Probabilistic Percolation (redMaPPer) catalog from \cite{Rykoff2014}, using an extended version and under the restriction of a minimum richness ($\lambda$) of 10, which in this redMaPPer catalog corresponds to masses $M_{\rm cluster} \gtrsim 0.8\times 10^{14} M_{\odot}$\footnote{It is important to note that, although lower levels of purity and completeness characterize samples from low-richness clusters \citep{Rykoff2014}, these factors are not expected to generate a spurious signal when comparing the properties of absorption-line systems in environments traced by such cluster-pairs with those in the field, assuming the presence of genuine inter-cluster filaments. Our methodology is inherently conservative, as any impurity or incompleteness would act to attenuate the observed signal, if any, rather than artificially enhance it. Specifically, spurious cluster detections could produce artificial filament identifications, which would dilute the filament-associated WHIM signal. Whereas catalog incompleteness could result in genuine filaments being classified as part of the field, thereby reducing the measured excess of WHIM tracers around filaments.}. With that constraint, our catalog was limited to \num{251093} galaxy cluster candidates. Although the redMaPPer detection algorithm is mainly based on photometry, around $50\%$ of the clusters have spectroscopic redshifts from their Brightest Cluster Galaxy (BCG). We use the BCG spectroscopic redshift as the cluster redshift ($z$) when available; otherwise, we use the photometric redshift.

The criteria that we impose to define our cluster-pairs catalog are:

\begin{itemize}
    \item The redshift of each cluster in a cluster-pair must be within $0.1  \leq z \leq 0.5$ (see Section~\ref{2.3} for the rationale behind the adopted limits).
    \item Both clusters must have a spectroscopic redshift.
    \item The redshift difference between the clusters of a pair must be  $< 2000$\,\kms.
    \item The transverse separation (projected in the sky) between two members of a cluster-pair must be $< 25$ Mpc.    
\end{itemize}

The third criterion was motivated to account for the typical velocity dispersion of galaxy clusters and its associated contribution to differences in cosmological redshift \citep{Tejos2016}. The final criterion was chosen based on theoretical results from N-body simulations in $\Lambda$CDM models, which indicate that beyond the selected threshold, the probability of filamentary structures between massive galaxy clusters drops below $\sim15\%$ (\citealt{Colberg2005}, \citealt{Gonzalez2010}).

Using these requirements, we obtained a total of \num{93296} cluster-pairs, which we cross-match with our QSO sample. 
\subsection{Selection of QSO sightlines}\label{2.2}
Our QSO catalog comes from FUV spectroscopy from the Hubble Space Telescope (HST) with the Cosmic Origins Spectrograph (COS) instrument. The QSOs were chosen from the combination of Sloan Digital Sky Survey Data Release 7 (SDSS DR7) \citep{Schneider2010}, the Baryon Oscillation Spectroscopic Survey (BOSS) \citep{Paris2012}, and the UV-bright Quasar Survey (UVQS) \citep{Monroe2016} intersecting the volume of SDSS, imposing a magnitude limit of FUV = $18.5$ AB. These comprise a large number of reliably identified QSOs with well-known spectroscopic redshifts that are bright enough for FUV spectroscopy in relatively short exposure times. These UV-bright QSOs were then cross-matched with our cluster-pairs, imposing the following criteria:

\begin{itemize}
    \item The redshift of the QSOs must be larger than the redshift of the cluster-pair probed, plus $3000$ \kms (rest-frame) from the QSOs redshift.
    \item The QSO sightline must be within an impact parameter of $3$ Mpc from the inter-cluster axis. 
\end{itemize}

\begin{figure}
   \centering
   \includegraphics[width=0.49\textwidth]{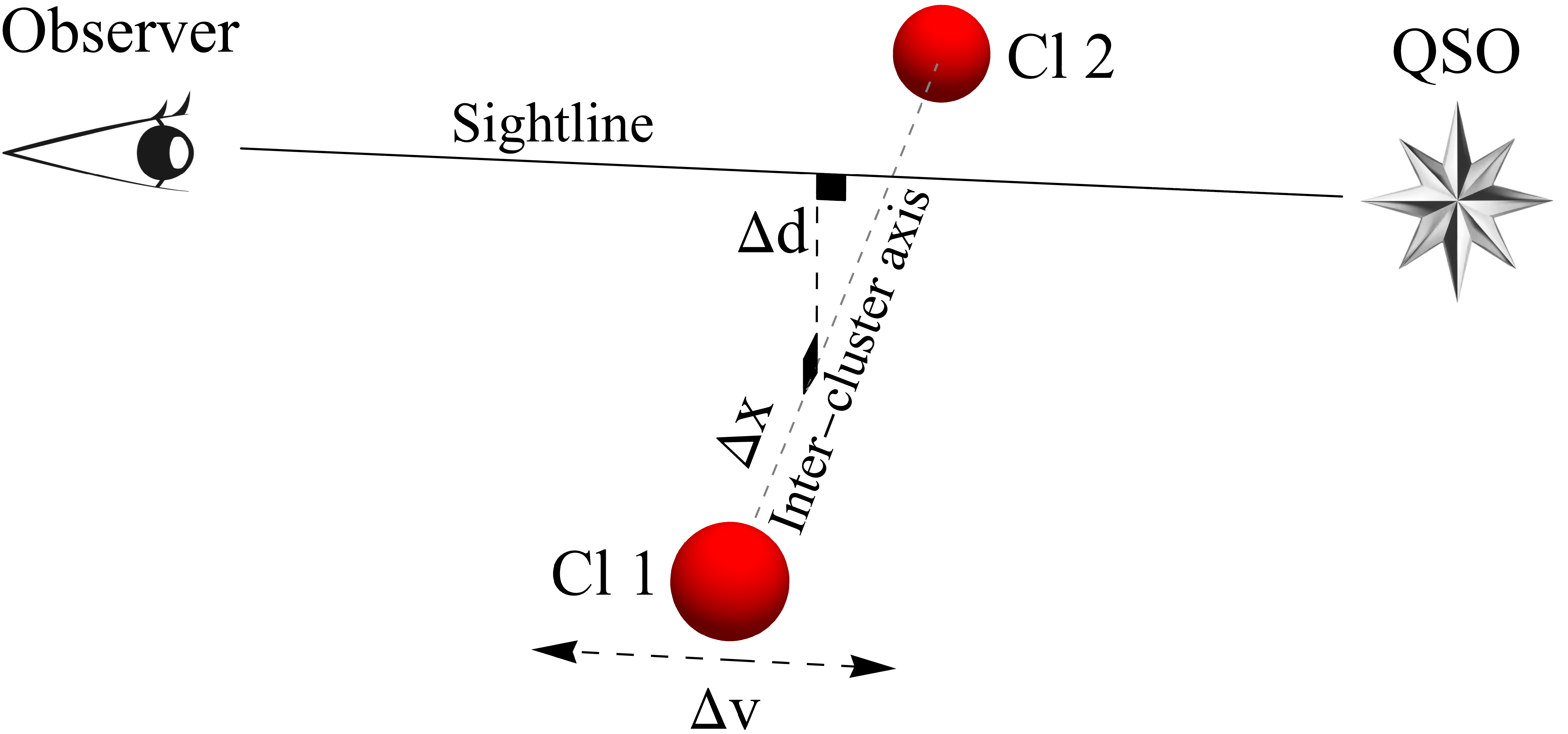}
      \caption{Schematic representation of our survey geometry, not shown to scale. A cluster-pair that meets our criteria is shown in red. A rest-frame velocity window around the position of the cluster-pair is defined as $\Delta v$; the distance between the QSO sightline and the inter-cluster axis is denoted by $\Delta d$; and the projected distance, along the inter-cluster axis, between the closest cluster and the QSO sightline is indicated as $\Delta x$.}
         \label{Fig:experiment}
   \end{figure}

The first criterion is imposed to avoid the proximity zone of the QSO, and the second criterion is imposed to maximize the probability of intersecting an inter-cluster filament. In Figure~\ref{Fig:experiment}, we show a schematic representation of the 3D geometry of our experiment. The inter-cluster axis is defined as the straight line connecting the two galaxy clusters in a pair, and the impact parameter, $\Delta d$, is defined as the minimum distance between the QSO sightline and the inter-cluster axis. Although this geometry is motivated by the expectation of having real inter-cluster filaments being probed by the QSO sightline, it is worth emphasizing that the underlying shape of the putative filaments is unknown. Therefore, the actual impact parameter might deviate from $\Delta d$; nevertheless, $\Delta d$ remains a reasonable estimate.

In this work, we use ten QSO sightlines that satisfy all these criteria observed from programs specifically designed for our study, namely HST GO 12958 and 15293 (PI Tejos)\footnote{We note that there are other QSOs in the Hubble Spectroscopic Legacy Archive (HSLA) catalog \citep{Peeples2017} that satisfy the criteria but these were not included in this work.}.
In Table~\ref{tab: QSOs_filament}, we summarize our QSO sample. 

\begin{table*}
\caption{Properties of the QSOs of the survey}
\label{tab: QSOs_filament} 
\centering
\begin{threeparttable}
\begin{tabular}{ccccccccc}
\hline\hline
ID$_{\text{QSO}}$ & QSO name & R.A. & Dec. & $z_{\text{QSO}}$ & Cluster-pair ID& \multicolumn{2}{c}{S/N$_{\text{QSO}}$}& HST program \\
 &  & (deg) & (deg) & & & \ion{H}{I} & \ion{O}{VI} & ID\\

(1) & (2) & (3) & (4) & (5)& (6) & \multicolumn{2}{c}{(7)} & (8)\\
\hline
QSO1 & J1410+2304 & 212.65999 & 23.07977 & 0.796 & 1-11 & 9 & 10 & 12958\tablefootmark{a}\\ 
QSO2 & J1456+2750 & 224.03606 & 27.83577 & 0.248 & 12-17 & 12 & 10 & 15293\\ 
QSO3 & J1257+4429 & 194.33150 & 44.49317 & 0.300 & 18-22 & 10 & 13 & 15293\\ 
QSO4 & J1410+0910 & 212.72264 & 9.17416 & 0.178 & 23-30 & 10 & 13 & 15293\\ 
QSO5 & J1419+3739 & 214.98633 & 37.65357 & 0.477 & 31-36 & 19 & 23 & 15293\\  
QSO6 & J1216+1819 & 184.13986 & 18.31866 & 0.449 & 37-43 & 10 & 11 & 15293\\  
QSO7 & J1358+0213 & 209.59995 & 2.22885 & 0.959 & 44-48 & 15 & 12 & 15293\\ 
QSO8 & J1210+0154 & 182.57646 & 1.90165 & 0.216 & 49 & 14 & 12 & 15293\\  
QSO9 & J1120+1104 & 170.08905 & 11.07633 & 0.495 & 50-53 & 12 & 8 & 15293\\ 
QSO10 & J1637+4254 & 249.44397 & 42.91379 & 0.544 & 54-61 & 10 & 9 & 15293\\ 
\hline
\end{tabular}
\end{threeparttable}
\tablefoot{(1) QSO ID. (2) Name of the QSO. (3-4) QSO coordinates. (5) Spectroscopic redshift of the QSO. (6) ID of the cluster-pairs (in Table~\ref{tab:cl-pairs}) associated with each QSO. (7) S/N per resolution element at the wavelength of the independent cluster-pairs in the QSO spectra for relevant absorption features (\ion{H}{I} and \ion{O}{VI}). (8) ID of the HST observational program.\\
\tablefoottext{a}{Published in \cite{Tejos2016}. }}
\end{table*}

The majority of the sightlines have an average $S/N$ greater than $10$ per resolution element, at the expected observed wavelength for relevant absorption features (\ion{H}{I} and \ion{O}{VI}). Figures~\ref{Fig:Survey_quality_appendix_HI} and~\ref{Fig:Survey_quality_appendix_OVI}  present the $S/N$ of the complete QSO sample from our survey, to examine the quality of our spectra. This constraint corresponds roughly to $3\sigma$ detection limits in column density of $10^{13}$ cm$^{-2}$ for \ion{H}{I} and $3\times10^{13}$ cm$^{-2}$ for \ion{O}{VI} (\citealt{Green2012}, \citealt{Butsky2019}) and is sufficient for independent constraints of the column density ($N$) and the Doppler parameter ($b$) via Voigt profile fitting. 

\subsection{Intervening cluster-pairs}

A total of $61$ intervening cluster-pairs were identified towards our $10$ QSO sightlines. According to previous studies, galaxy clusters themselves cluster together; therefore, some cluster-pairs should reside within the same physical structures (\citealt{White1991}, \citealt{Benson2001}). For a given QSO sightline, we grouped cluster-pairs that are separated by less than $1000$ \kms (rest-frame), as these systems are likely to trace the same filamentary structures, if any \citep{Tejos2016}. We therefore treated each group of associated cluster-pairs as a single independent system. The velocity value was adopted to account for the velocity dispersion typical of galaxy clusters and the associated contribution to differences in cosmological redshift. As representatives of independent cluster-pairs, we selected the pair with the smallest impact parameter $\Delta d$. Accordingly, the $61$ cluster-pairs associated with their respective QSO sightlines were grouped into $22$ independent systems. Table~\ref{tab:cl-pairs} presents the galaxy cluster-pairs associated with their respective QSO sightlines.

\subsection{HST/COS observations}
\label{2.3}

The COS instrument, in its FUV mode, is a medium-resolution spectrograph ($\lambda/\Delta \lambda \approx 16000-21000$ or FWHM $\approx 0.07-0.09$ \AA) with wavelength coverage of 1135–1800 \AA. We employed the G130M and G160M gratings, which are the two moderate-resolution FUV channels of COS. This grating combination is sensitive to \ion{H}{I} absorption through Ly$\alpha$ at $z \lesssim  0.47$ and through Ly$\beta$ and higher-order Lyman transitions at $0.10\lesssim z \lesssim 0.90$ \citep{Danforth2016}. The G130M grating is sensitive to detect the $1032$ \AA\, transition -the strongest of \ion{O}{VI}- over the redshift range $0.10 \lesssim z \lesssim 0.40$, whereas the G160M grating is sensitive to $0.30 \lesssim z \lesssim 0.70$ (\citealt{Kerman2023}, \citealt{Koplitz2024}). To trace \ion{H}{I} and \ion{O}{VI} simultaneously at low redshift, we constrain our analysis to $0.1\leq z\leq0.5$. 

\subsection{QSOs data processing}

We process the QSO spectra in the same fashion as \cite{Pessa2025}, briefly described as follows. First, we retrieved the individual exposures of our QSO spectra from the Mikulski Archive for Space Telescopes (MAST). We downloaded the reduced files {\tt x1d.fits}, which had been previously uniformly processed by the HST/CalCOS pipeline. CalCOS calibrates COS data by rectifying instrumental effects, producing a wavelength calibration tailored to each exposure, and finally extracting a one-dimensional spectrum calibrated for flux, for the entire observation \citep{James2022}. Nevertheless, we inspected the 2D spectral files to assess their quality.
Subsequently, we used \linetools\footnote{\url{https://github.com/linetools/linetools}.} \citep{Prochaska2016} to rebin the original spectra into a coarser single linear wavelength scale of $0.0395$\,\AA\,pixel$^{-1}$, which corresponds roughly to two pixels per resolution element \citep{Tejos2016}. In spectral regions covered by both gratings, we average the spectra, ensuring flux conservation during rebinning. The calibrated one-dimensional final spectra were further analyzed using the {\sc linetools} {\tt lt\_continuumfit} script to fit a global pseudo-continuum to it\footnote{i.e., including intrinsic broad emission lines and Galaxy’s damped Ly$\alpha$ system wings \citep{Tejos2016}.}.

\section{Analysis}
\label{analysis}

We search for intervening absorption lines in each QSO spectrum and characterize their observational parameters via Voigt profile fitting. We focus our analysis on total \ion{H}{I}, NLAs, BLAs, and \ion{O}{VI} absorption lines. Hereafter, in this work, total \ion{H}{I} refers to the combined contribution of both BLAs and NLAs.

\subsection{Spectral lines identification}

We identify all absorption lines in each QSO sightline using the software IGM-Guesses\footnote{\url{https://pyigm.readthedocs.io/en/latest/igmguesses.html}.}, not limited to the wavelength regions corresponding to the redshifts of known structures. This software is a graphical user interface (GUI) available as part of the {\sc pyigm} package\footnote{\url{https://github.com/pyigm/pyigm}.} and facilitates the search for and identification of absorption features in the spectra of background sources. However, IGM-Guesses only produces initial guesses for the redshift, column density, and Doppler parameter for subsequent Voigt profile fitting. To characterize all absorption lines, we followed the algorithm described in \cite{Tejos2016} (see also \citealt{Holguin2024} and \citealt{Pessa2025}), which is summarized as follows.

First, we must specify a confidence category for identifications, following the recommendations in the IGM-Guesses documentation. We do this through qualitative categories that indicate the reliability of the detection. Category `a' (reliable) is assigned to the detections with multiple transitions of a given ion where at least two of them are available and clearly visible in the spectrum. Category `b' (possible) is assigned when we detect absorption components from single transition ions that are at the same redshift (within a reasonable velocity window) as another reliable component, since they exhibit similar kinematic structure, which increases the reliability of the detection, or when ion components have several possible transitions, but for some reason, only one of them is clearly detected. Category `c' (uncertain) is initially assigned when components, based on the user experience, are likely to be misidentified or to exhibit an unknown feature. Additionally, we imposed a detection significance criterion based on the equivalent width ($W_r$) and its uncertainty ($\delta W_r$). Absorbers with reliability `b' with $W_r/\delta W_r < 3$ are reassigned to reliability `c'.

With these criteria, we use the 5-step IGM-Guesses procedure \citep{Tejos2016}:

\begin{enumerate}
    \item Step 1: We identify all absorption components available at redshift $z=0$ with a velocity window of $\pm500$ \kms in order to ensure the identification of the lines from the interstellar medium (ISM) of the Milky Way and/or Local Group Systems.
    \item Step 2: This step replicates Step 1, but at the redshift of the QSO ($z_{QSO}$) with a velocity window of $\pm1500$ \kms to ensure the identification of the lines associated with the QSO environment.
    \item Step 3: We search for absorption lines from the IGM. To do so, we identify \ion{H}{I} with at least two transitions, starting at $z=z_{QSO}$ and scanning down to $z=0$, and assign them to the reliability ‘a’. This step involves identifying the entire Lyman series covered by the spectrum of a given component.
    \item Step 4: We search for metal lines in the well-identified \ion{H}{I} absorption detected in the previous step. When the wavelength coverage allows detection of multiple transitions of a single ion, we require that their relative positions coincide; in cases of multiple adjacent components blending, we require them to have consistent kinematic structures across multiple transitions of the same ion.
    \item Step 5: We assume all unidentified absorption features to be \ion{H}{I} Ly$\alpha$ starting from $z=z_{QSO}$ down to $z=0$, and assign them to the ‘b’ category. Then we repeat the previous step, and if metal lines are found, we re-assign the corresponding identification and reliability, and categorize the \ion{H}{I} line as category ‘a’. 
\end{enumerate}
Finally, we save these absorption features and the initial guesses on ($z$, $N$, $b$).

\subsection{Observational parameters estimation}

We use the {\sc Veeper}\footnote{\url{https://github.com/jnburchett/veeper}} Voigt profile fitting software based on our initial guesses described above. Veeper is a collection of tools for generating and fitting Voigt profiles of absorption lines, particularly in QSO spectra, accounting for the non-Gaussian line spread function (LSF) of COS. We fit Voigt profiles to all absorption lines in the QSO spectrum, yielding the converged values for $z$, $\log N$, and $b$, along with their uncertainties. 
Additionally, we calculated the rest-frame equivalent width with {\sc Linetools}, and its uncertainty was calculated as in \cite{Tejos2016} (see their Section 4.3).

In our analysis of ten QSO sightlines, we detected 99, 83 and 141 NLAs with reliabilities `a', `b' and `c’, respectively; 11, 76 and 66 BLAs with reliabilities `a', `b' and `c’, respectively; and 34, 6 and 4 \ion{O}{VI} with reliabilities `a', `b' and `c’, respectively. However, in this work, we restrict our analysis only to absorption features with reliabilities `a' and `b’. Therefore, the subsequent analysis should be regarded as providing lower limits, as the most uncertain absorbers -although potentially genuine- may have been excluded.

The observational parameters derived from the absorption lines detected and fitted across all QSO sightlines in our survey are included in the Supplementary Material. As an example, the table of observational parameters for a subset of the absorption components is shown in Table~\ref{tab: parametros} in the appendix.

\begin{figure*}[ht]
  \includegraphics[width=\textwidth]{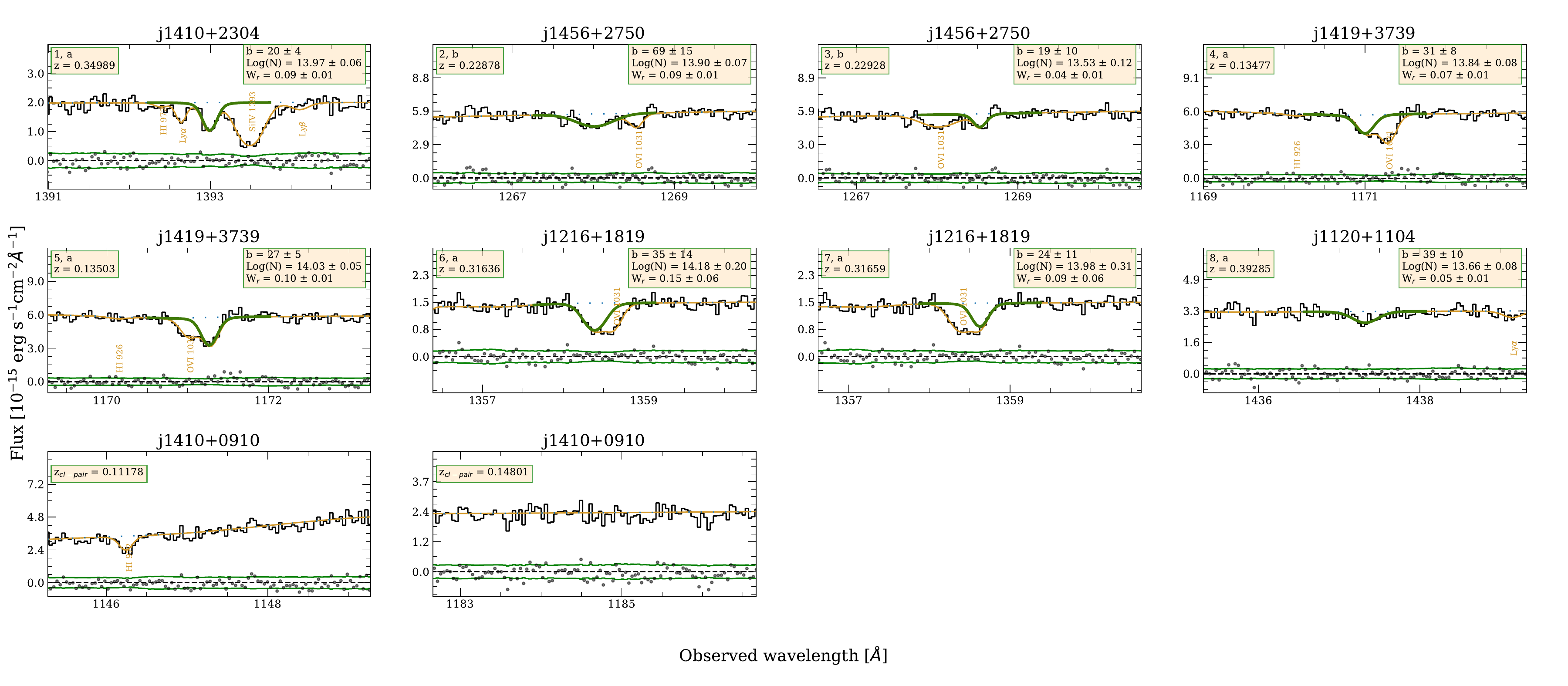}
  \caption{Identified \ion{O}{VI} in a velocity window of $\Delta v=\pm1000$\,\kms and at impact parameter $\Delta d\leq 3$ Mpc from the cluster-pairs. The panels show the best-fitting Voigt profile for each \ion{O}{VI} (thick \ovicolor line) and the full QSO spectrum model (thin gold line). The dotted blue line shows the continuum level. The additional transitions identified in the same wavelength range are labeled in gold. At the top-right corner of each panel, we show the derived observational parameters: column density $\log(N/cm^{-2})$, Doppler parameter in \kms, and equivalent width in \AA. In the top-left corner of each panel, we show the ID, reliability, and redshift of each absorption identified. Each plot title specifies the QSO to which the absorption feature corresponds. The last two panels show the spectrum of QSO J1410+0910 at the redshifts of the cluster-pairs, where no transitions are detected. The corresponding plot for the $1037$ \AA\, transition of \ion{O}{VI} is shown in Figure~\ref{Fig:Survey_OVI_transition} in the appendix.}
  \label{Fig:Survey_panels_OVI}
\end{figure*}

In Figures~\ref{Fig:Survey_panels_NLA}, ~\ref{Fig:Survey_panels_BLA}, and~\ref{Fig:Survey_panels_OVI}, we show the NLAs (in \nlacolor), BLAs (in \blacolor), and \ion{O}{VI} (in \ovicolor) detections, respectively, that we found at a rest-frame velocity window of $\Delta v=\pm 1000$ \kms and at impact parameters $\Delta d\leq 3$ Mpc from the cluster-pairs axes. We adopt these fiducial values for comparison with \cite{Tejos2016}, which uses identical values. The fiducial value for $\Delta v$ is consistent with the third constraint from Section~\ref{2.1}. The selection of the fiducial value for $\Delta d$ is imposed as part of the experimental design (see Section~\ref{2.2}).

\begin{figure*}[ht]
\includegraphics[width=\textwidth]{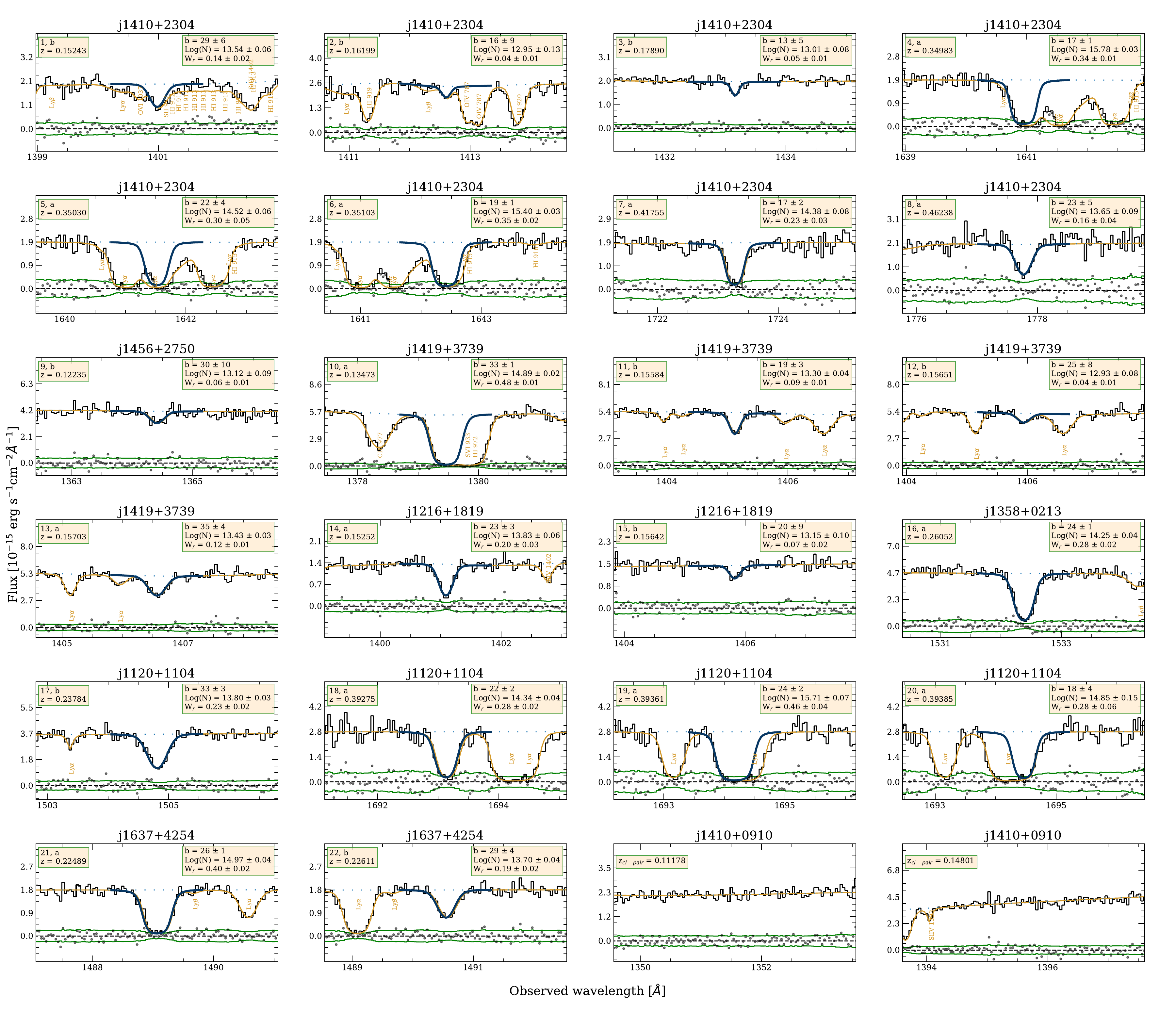}
  \caption{Identified NLAs in a velocity window of $\Delta v=\pm1000$ \kms and at impact parameter $\Delta d\leq 3$ Mpc from the cluster-pairs. The panels show the best-fitting Voigt profile for each NLA (thick \nlacolor line) and the full model QSO spectrum (thin gold line). The dotted blue line shows the continuum level. The additional transitions identified in the same wavelength range are labeled in gold. At the top-right corner of each panel, we show the derived observational parameters: column density $\log(N/cm^{-2})$, Doppler parameter in \kms, and equivalent width in \AA. In the top-left corner of each panel, we show the ID, reliability, and redshift of each absorption identified. Each plot title specifies the QSO to which the absorption feature corresponds to. The last two panels show the spectrum of QSO J1410+0910 at the redshifts of the cluster-pairs, where no transitions are detected.}
  \label{Fig:Survey_panels_NLA}
\end{figure*}

\begin{figure*}[ht]
  \includegraphics[width=\textwidth]{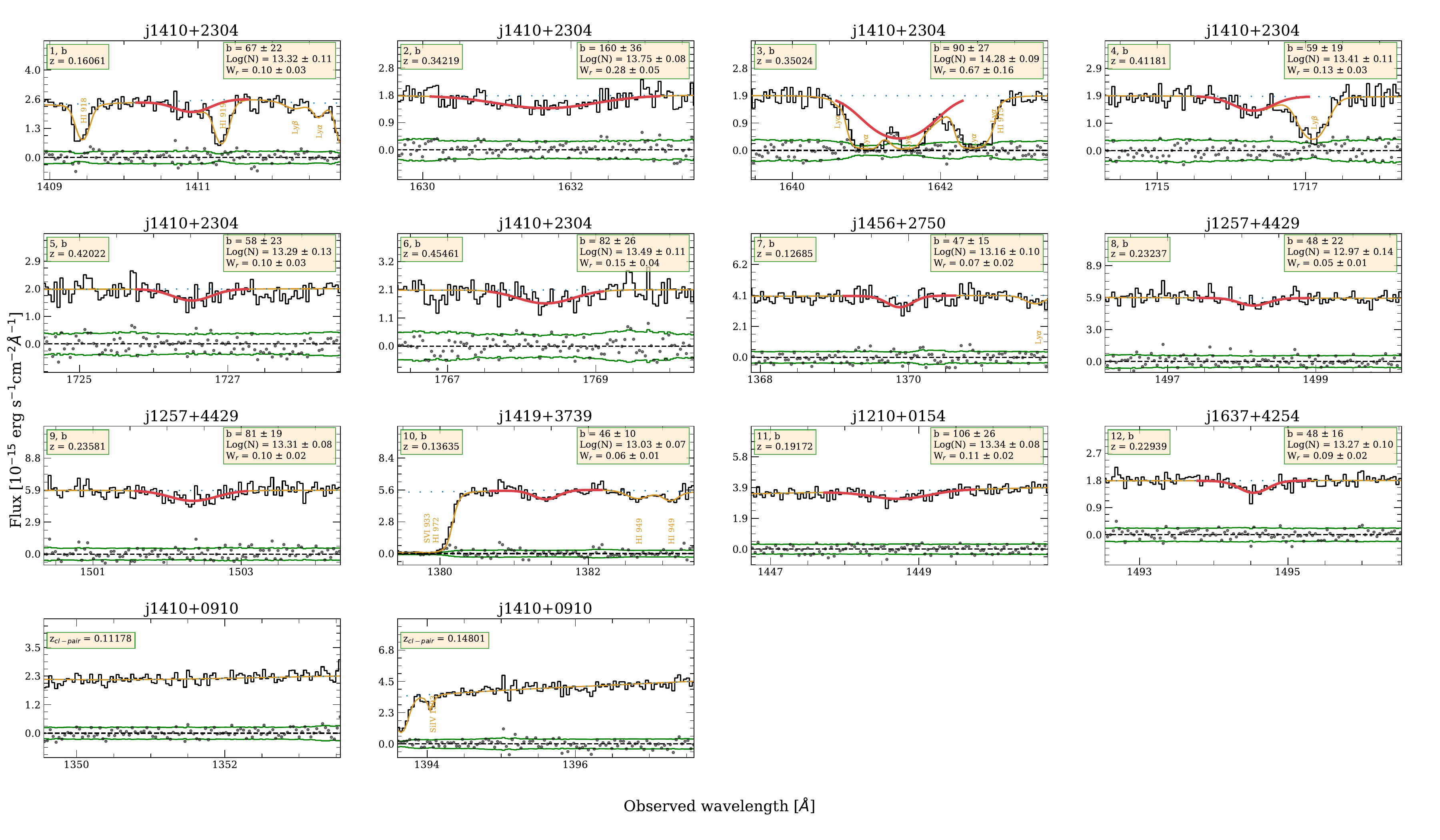}
  \caption{Identified BLAs in a velocity window of $\Delta v=\pm1000$ \kms and at impact parameter $\Delta d\leq 3$ Mpc from the cluster-pairs. The panels show the best-fitting Voigt profile for each BLA (thick \blacolor line) and the full QSO spectrum model (thin gold line). The dotted blue line shows the continuum level. The additional transitions identified in the same wavelength range are labeled in gold. At the top-right corner of each panel, we show the derived observational parameters: column density $\log(N/cm^{-2})$, Doppler parameter in \kms, and equivalent width in \AA. In the top-left corner of each panel, we show the ID, reliability, and redshift of each absorption identified. Each plot title specifies the QSO to which the absorption feature corresponds. The last two panels show the spectrum of QSO J1410+0910 at the redshifts of the cluster-pairs, where no transitions are detected.}
  \label{Fig:Survey_panels_BLA}
\end{figure*}

\subsection{Redshift path sensitivity function}
\label{sect_condition_deltaz}
The redshift path of a QSO is governed by the observed wavelength range, which is instrument dependent, the $S/N$, and the rest-frame wavelength of the transition representing the type of system being surveyed. Given that our survey compiles absorber path lengths across several quasar sightlines, we needed to estimate the redshift path sensitivity function $g(z)$, which quantifies the effective redshift path over which a survey is sensitive to detect a given absorption line. For our survey of 10 QSO sightlines, we computed $g(z)$ to determine the feasibility of detecting \ion{H}{I} and \ion{O}{VI} absorption features as a function of redshift. The function $g(z)$ is defined according to Equation~\eqref{Eq:redshift path survey} \citep{Churchill2025}:
\begin{equation}
    g(z)=\sum_{k=1}^{M} H(z-z_{k}^{min})H(z_{k}^{max}-z),
    \label{Eq:redshift path survey}
\end{equation}
here, the sum is over all QSO sightlines, $M$ is number of QSOs in the survey, the $k$-th QSO has a redshift path ranging from $z_{k}^{min}$ to $z_{k}^{max}$, and $H(t)$ is the Heaviside step function.
The redshift path ($\Delta z$) is calculated using the constraints given in Appendix~C of \cite{Tejos2016}. In summary, the $\Delta z$ $(\Delta d, \Delta v, W_{r}^{min})$ was estimated for a maximum rest-frame velocity difference within $\Delta v$ from its cluster-pair, a maximum $\Delta d$ from the QSO sightline, and a minimum $W_{r}^{min}$ for the detections. We considered only regions where the QSO spectrum covers the redshift range of $0.1\leq z\leq 0.5$ for the absorption of interest (\ion{H}{I} and \ion{O}{VI}). We mask out regions with fluxes below 50\% of the value of the continuum fit; this is because of biases against detecting absorption systems on top of strong absorption lines. We also masked out regions within $\pm200$ \kms from expected Galactic absorbers. After that, we estimated the minimum rest-frame equivalent width to observe a transition at rest-frame wavelength, $\lambda_{0}$, along the spectrum by this relation $W_{r,\lambda_{0}}^{min} = 3\frac{\lambda_{0} R}{\langle S/N \rangle}$, with R the resolution of the COS spectrograph (taken to be $R = 20000$) and $\langle S/N \rangle$ is the signal-to-noise ratio per resolution element of the spectrum at $\lambda_{obs}=\lambda_0(1+z)$. 
We then identified portions of the spectrum satisfying the criteria of $W_{r,\lambda_{0}}^{min} \geq W_{r}^{min}$. Using these constraints and Equation~\eqref{Eq:redshift path survey}, we obtained the redshift path sensitivity functions, shown in Figure~\ref{Fig:g_z_HI_OVI} in the appendix.

The total redshift path ($\Delta z$) of the survey is calculated by adding the contribution of the redshift path of each species from each QSO, according to Equation~\eqref{Eq:redshift_path}.
\begin{equation}
\Delta z = \int_{0}^{\infty}  g(z) \mathrm{d}z,
    \label{Eq:redshift_path}
\end{equation}
\noindent where $\mathrm{d} z = \Delta \lambda_{pix} /\lambda_{r}$, is the redshift value corresponding to a single pixel, $\Delta \lambda_{pix}=0.0395$ \AA, and $\lambda_{r}$ is the rest-frame wavelength of the transition representing the system being surveyed, in our case \ion{H}{I} ($\lambda_{r} = 1215.67$ \AA) and \ion{O}{VI} ($\lambda_{r} = 1031.9261$ \AA).  

\begin{figure}
     \centering
     \begin{subfigure}[b]{0.49\textwidth}
         \centering
         \includegraphics[width=\textwidth]{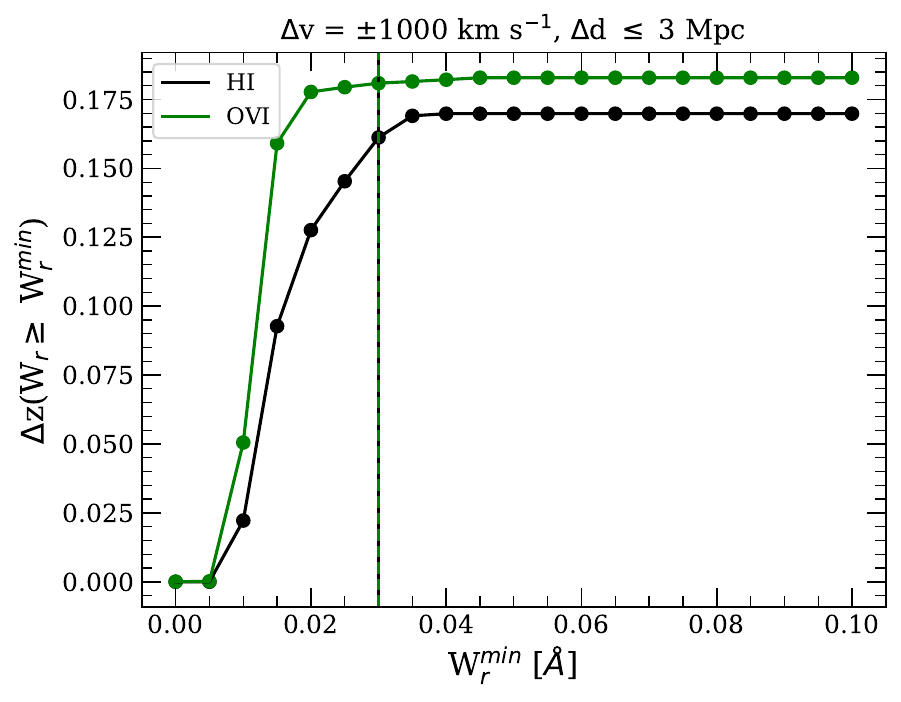}
     \end{subfigure}
     
     \begin{subfigure}[b]{0.49\textwidth}
         \includegraphics[width=\textwidth]{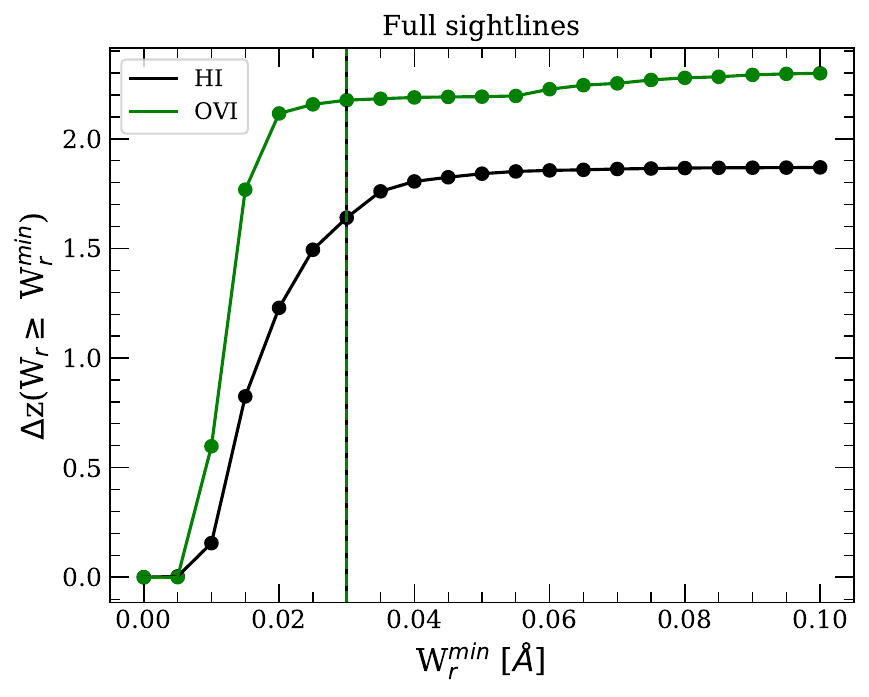}
     \end{subfigure}
        \caption{Redshift path as a function of minimum rest-frame equivalent width for the survey. In the top panel, we show the corresponding redshift path associated with regions of our sample spectra with rest-frame velocity differences of $\Delta v =\pm 1000$ \kms from their cluster-pair at impact parameters smaller than $\Delta d = 3$ Mpc. In the bottom panel, we show the redshift path, $\Delta z$, across the entire sample for \ion{H}{I} in black and \ion{O}{VI} in green, shown as a function of the minimum rest-frame equivalent width, $W_{r}^{min}$. The vertical lines indicate the minimum rest-frame equivalent width adopted in our analysis for \ion{H}{I} and \ion{O}{VI}, namely $0.03$ \AA\ in both cases. }
        \label{Fig:Redshift_path_full_survey}
\end{figure}

The top panel of Figure~\ref{Fig:Redshift_path_full_survey} shows the total redshift path corresponding to regions of our survey spectra within rest-frame velocity differences within $\Delta v =\pm 1000$ \kms from their cluster-pairs at impact parameter $\Delta d \leq 3$ Mpc. In the bottom panel, we show the total redshift path of the survey across the entire sample for detections of \ion{H}{I} and \ion{O}{VI}. The redshift path is calculated at different rest-frame equivalent width limits, $W_{r}^{min}$. This function indicates the completeness of a survey by providing a measure of the available search path as a function of the strength of absorbers that could be detected (\citealt{Churchill1999}, \citealt{Narayanan2007}, \citealt{Tejos2016}). As expected, weak lines (small $W_{r}^{min}$) can be reliably detected only in the highest signal-to-noise regions of the spectra, whereas strong lines (large $W_{r}^{min}$) can be surveyed reliably across nearly the entire available redshift path. The similar shapes of the upper and lower curves for \ion{H}{I}, as well as those for \ion{O}{VI}, in Figure~\ref{Fig:Redshift_path_full_survey} suggest comparable levels of completeness between the regions surrounding cluster-pairs and the full spectra for each tracer. The vertical lines indicate the minimum rest-frame equivalent width adopted for the subsequent analysis for \ion{H}{I} and \ion{O}{VI}, namely $0.03$ \AA\ in both cases, which corresponds to high completeness levels, above 88\% for both absorbers at both spatial samples. The targeted redshift path of our survey at the selected minimum equivalent width is $1.64$ for \ion{H}{I} and $2.18$ for \ion{O}{VI}.

\begin{figure*}
     \begin{subfigure}[b]{0.49\textwidth}
         \centering
         \includegraphics[width=0.9\textwidth]{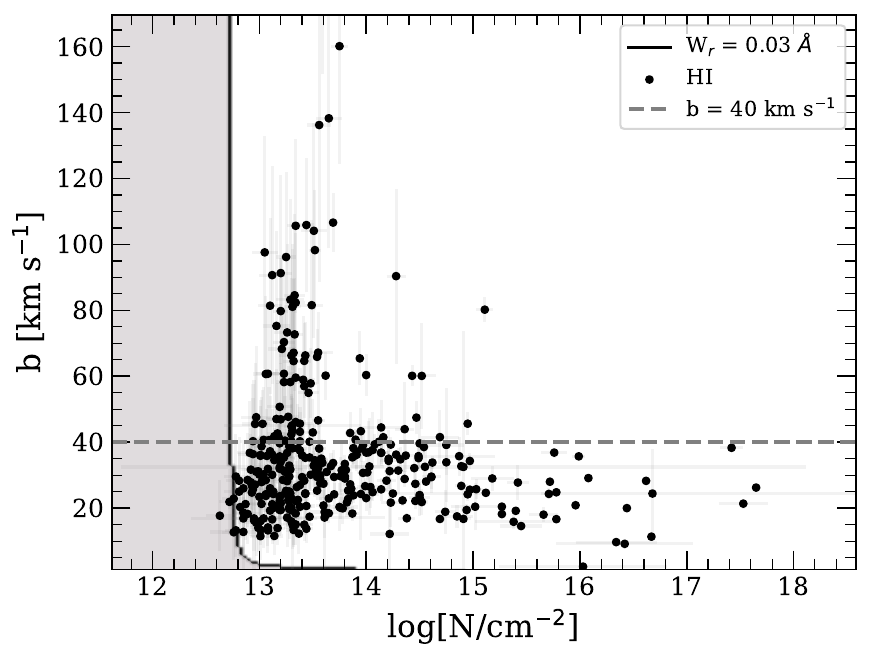}
     \end{subfigure}
     \hfill
     \begin{subfigure}[b]{0.49\textwidth}
    \includegraphics[width=0.9\textwidth]{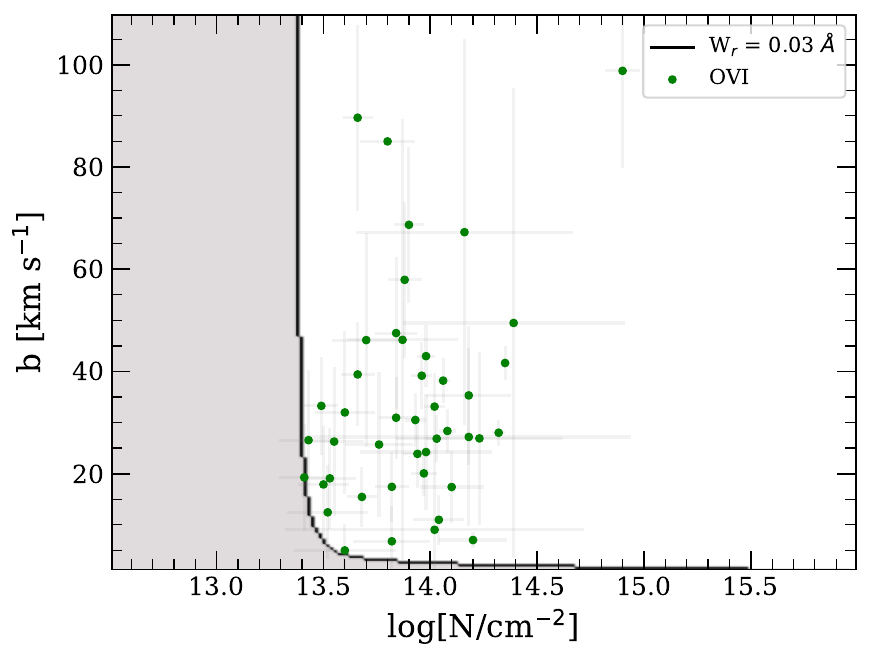}
     \end{subfigure}
        \caption{ Doppler parameter distribution as a function of the column densities for our sample of \ion{H}{I} in the left panel and \ion{O}{VI} in the right panel. The uncertainties are shown in the background, and the gray dashed horizontal line represents our limit in the Doppler parameter of $40$ \kms\ which defines BLAs. The shaded area corresponds to W$_{r}<0.03$ \AA, indicating the region of our survey where we have low completeness (see section \ref{sect_condition_deltaz}).}      
        \label{Fig:b_logn_HI_OVI}
\end{figure*}

In Figure~\ref{Fig:b_logn_HI_OVI}, we show the Doppler parameter distribution as a function of the column densities, for our sample of \ion{H}{I} (left panel) and \ion{O}{VI} (right panel), between $0.1 \leq z  \leq  0.5$. The uncertainties are shown in the background, and the gray dashed horizontal line represents our limit in the Doppler parameter of $40$ \kms, which we use to split BLA in the upper part of the \ion{H}{I} plot and NLA samples in the bottom part. The shaded area corresponds to regions with W$_{r}<0.03$ \AA, indicating the area of our survey with low completeness. 

We observe that for NLAs, there are many detections near the contour, which limits the detection capabilities of our survey, whereas for BLAs, there are fewer detections near that contour. This behavior is likely consistent with the fact that there are more NLA than BLA detections and that $S/N$ limitations affect BLAs more than NLAs, for a fixed W$_r$. Additionally, the few \ion{H}{I} absorbers that are located in the shaded region correspond to QSO J1419+3739, which has a high S/N (see Figure~\ref{Fig:Survey_quality_appendix_HI} in the appendix), allowing us to detect those absorbers. We note that the contour is defined theoretically (without S/N consideration), whereas the adopted limit is observationally motivated (see Section \ref{sect_condition_deltaz}).

\subsection{Redshift number density around cluster-pairs}

We estimated the redshift number density of absorption lines ($d\mathcal{N}/dz$), which is the number of absorbers detected for a given ion species per unit redshift. We used this to characterize inter-cluster filaments as a function of a rest-frame velocity window ($\Delta v$) around the redshift of the 22 independent cluster-pairs. Following \cite{Tejos2016}, to estimate the redshift number density, we counted $\mathcal{N}(\Delta d, \Delta v, W_{r}^{min})$ as the number of absorption components found within $\Delta v$ from the identified cluster-pair, having rest-frame equivalent widths $W_{r} \geq W_{r}^{min}$, and considering only cluster-pairs with a distance between the inter-cluster axis and QSO sightline smaller than $\Delta d$. Then the redshift number density is calculated by Equation~\eqref{Eq:redshift_number_density}:
\begin{equation}
    \frac{\mathrm{d} \mathcal{N}}{\mathrm{d} z}(\Delta d, \Delta v, W_{r}^{min})=\frac{\mathcal{N} (\Delta d, \Delta v, W_{r}^{min})}{\Delta z (\Delta d, \Delta v, W_{r}^{min})},
    \label{Eq:redshift_number_density}
\end{equation}
where $\Delta z (\Delta d, \Delta v, W_{r}^{min})$ is the redshift path of \ion{H}{I} or \ion{O}{VI} also as a function of these thresholds. To calculate the redshift number density of NLAs and BLAs, we use the redshift path of \ion{H}{I}.

The uncertainty in the redshift number density is dominated by the Poissonian uncertainty in $\mathcal{N}$\footnote{For low-number statistics, we approximate the upper ($\sigma_\mathcal{N}^+$) and lower ($\sigma_\mathcal{N}^-$) uncertainties in the number of absorption components by Equation~\eqref{Eq:sigma_dN/dz} \citep{Gehrels1986} at the 1-$\sigma$ confidence level:
\begin{equation}
    \sigma _{\mathcal{N}}^{+}\approx \sqrt{\mathcal{N}+3/4}+1\\
    \sigma _{\mathcal{N}}^{-}\approx \sqrt{\mathcal{N}-1/4}
    \label{Eq:sigma_dN/dz}
\end{equation} 
We note that the approach employed to determine the uncertainties yields a robust estimate, as it agrees with the Jackknife variance (\citealt{Wolter2007}, \citealt{Tejos2014}).} \citep{Tejos2016}.

The redshift number density of the field ($d\mathcal{N}/dz|_{field}$) characterizes the specific absorption incidence rate of a given ion species in the IGM and is determined from randomly selected sightlines. This value allows for a comparison of $d\mathcal{N} /dz$ for a particular ion species with the typical level of absorption expected in the field for that same ion species. 
To estimate the expected field value for our survey, we used Equation~\eqref{eq:my_expected_field}:
\begin{equation}
    \dfrac{d\mathcal{N}}{dz}\Bigg|_{field}\approx \frac{\mathcal{N}_{tot}}{\Delta z_{tot}},     \label{eq:my_expected_field}
\end{equation}
where $\mathcal{N}_{tot}$ is the total number of absorption lines of the analyzed absorber in the survey, restricted to the range $0.1 \leq z \leq 0.5$, and $\Delta z_{\mathrm{tot}}$ is the total redshift path, also at those redshift limits. The total redshift path does not exclude regions surrounding cluster-pairs\footnote{Because the survey sightlines were selected to probe filaments, the expected field could alternatively be calculated by excluding regions near cluster-pairs. Nevertheless, our current estimations of the expected field remain consistent with established literature \citep{Danforth2016}. Furthermore, an analysis excluding these cluster-adjacent regions yields values that are consistent with the reported results within the calculated uncertainties. Notably, such an exclusion would reduce the expected field value, thereby amplifying the observed relative excesses, particularly for BLAs and \ion{O}{VI}, and further reinforcing the primary conclusions of this study.}, and is $\sim10$ times larger than the redshift path associated with independent cluster-pairs at the fiducial values. The estimation of the statistical uncertainty in the expected field is also taken only from the contribution of the Poissonian uncertainty of $\mathcal{N}_{tot}$ using Equation~\eqref{Eq:sigma_dN/dz}.

For the redshift number density of the field, we compare our results, as given by Equation~\eqref{eq:my_expected_field}, with estimations derived from \cite{Danforth2016}, $d\mathcal{N}/dz|_{field_D}$. For \ion{H}{I}, we employ a survey of \cite{Danforth2016} consisting of $82$ UV-bright QSOs, and the expected field was calculated with Equation~\eqref{Eq:dNdz_field}:
\begin{equation}
     \frac{d\mathcal{N}(>N)}{dz}\Bigg|_{field}=C_{14}\left ( \frac{N}{10^{14}cm^{-2}} \right ) ^{-(\beta-1)},
     \label{Eq:dNdz_field}
\end{equation}
where $N$ is the average column density of the lowest 5\% absorbers of a specific ion species (the average was taken to avoid outliers). We consider absorbers with a significance level (SL) $\geq10$ \citep{Danforth2016} and with a fit concern of 0, indicating that there is no doubt the detection corresponds to the specific identified ion species \citep{Danforth2016}. 
The normalization constant $C_{14}$ and $\beta$ are fitting parameters. 
The best-fitting values for these constants have been reported in the literature for studies of differential distribution of absorbers for $12\leq\log (N/cm^{-2})\leq17$. For \ion{H}{I}, the reported values are $\beta=1.67\pm0.01$, and $C_{14}=23\pm1$ \citep{Danforth2016}. The uncertainties are calculated using error propagation of the uncertainties in the fit parameters following Equation~\eqref{Eq:dNdz_field}. For \ion{O}{VI}, the expected field is derived directly from \cite{Danforth2016}. Since \cite{Danforth2016} does not provide explicit values for the fitting parameters to calculate the redshift number density of the field, we estimate it using the $\log N$ ranges, the number of \ion{O}{VI} absorbers identified, $\mathcal{N}_{OVI}$, and the redshift path for \ion{O}{VI}, $\Delta z_{OVI}$, reported in their Table~5. Within the range of $\log N$ adopted in this work, using the corresponding values of $\mathcal{N}_{OVI}$, and $\Delta z_{OVI}$, we compute the redshift number density of \ion{O}{VI} according to  $\frac{d\mathcal{N}(>N)}{dz}\Bigg|_{field_{D}}=\sum_{j=1}^{\mathcal{N}_{OVI}}\frac{1}{\Delta z_{\text{\ion{O}{VI}, }j}}$, and we estimate the uncertainties using $ \sigma^{2}=\sum_{j=1}^{\mathcal{N}_{OVI}}\frac{1}{\Delta z_{\text{\ion{O}{VI}, }j}^{2}}$ \citep{Churchill2025}.

\section{Results}
\label{results}

We obtained the redshift number density of our survey for the total \ion{H}{I}, NLA, BLA, and \ion{O}{VI} components around cluster-pairs and determined the relative excesses of these samples with respect to their respective expected field values. 

\subsection{Redshift number density of the survey}\label{sec:4.1}
\begin{figure*}
 \centering
 \begin{subfigure}[ht]{0.49\textwidth}
     \centering
     \includegraphics[width=\textwidth]{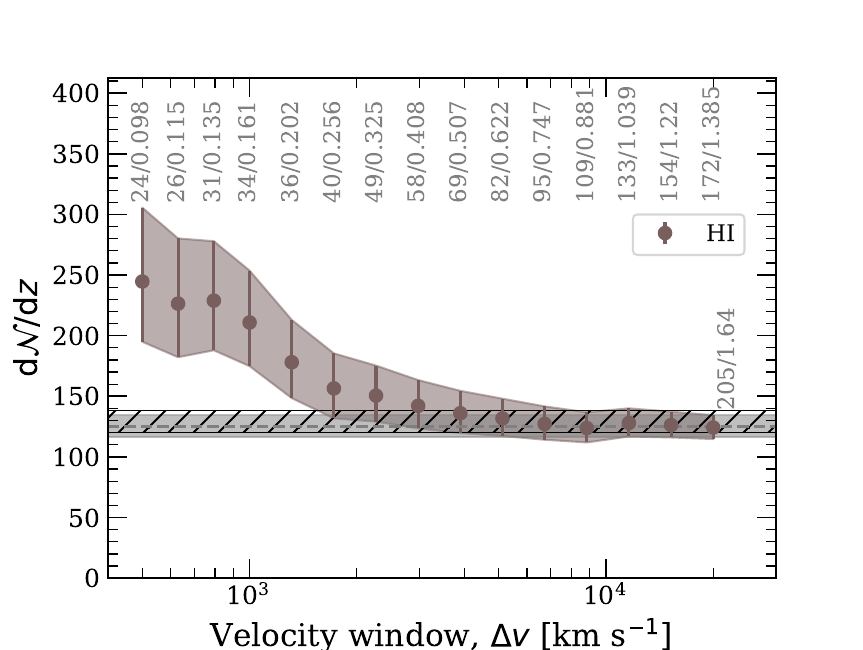}
 \end{subfigure}
 \hfill
 \begin{subfigure}[ht]{0.49\textwidth}
     \centering
     \includegraphics[width=\textwidth]{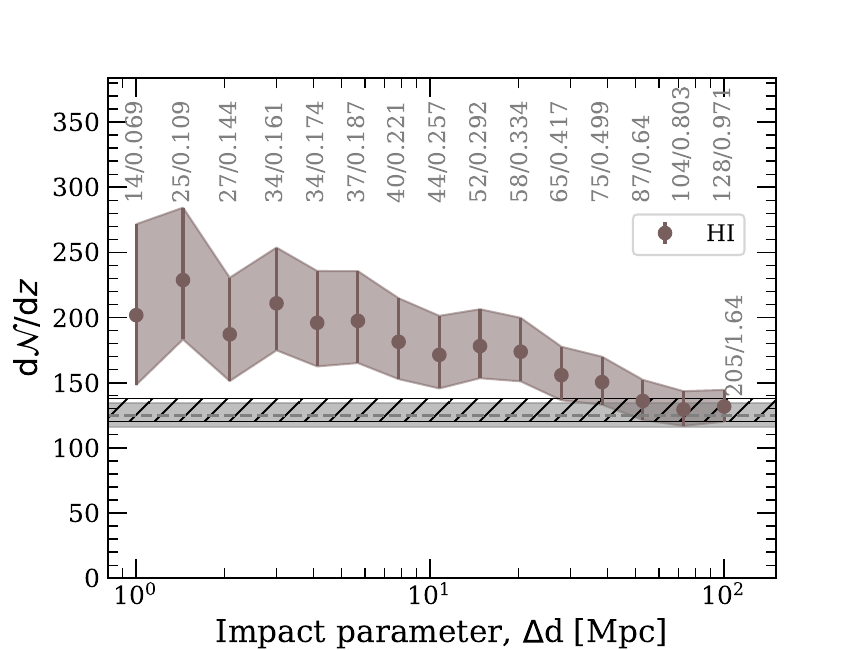}
 \end{subfigure}
 \hfill
 \begin{subfigure}[ht]{0.49\textwidth}
     \centering
     \includegraphics[width=\textwidth]{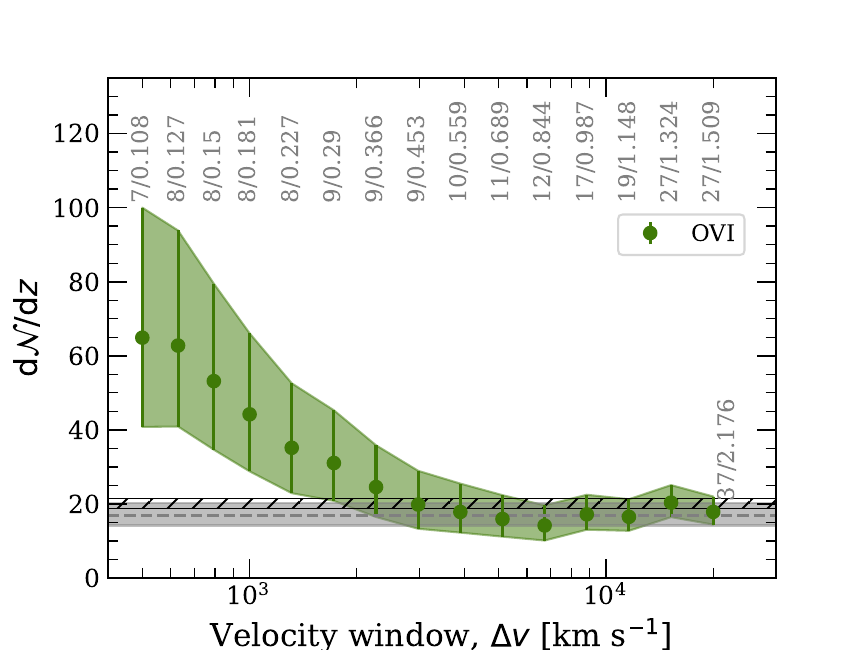}
 \end{subfigure}
 \hfill
 \begin{subfigure}[ht]{0.49\textwidth}
     \centering
     \includegraphics[width=\textwidth]{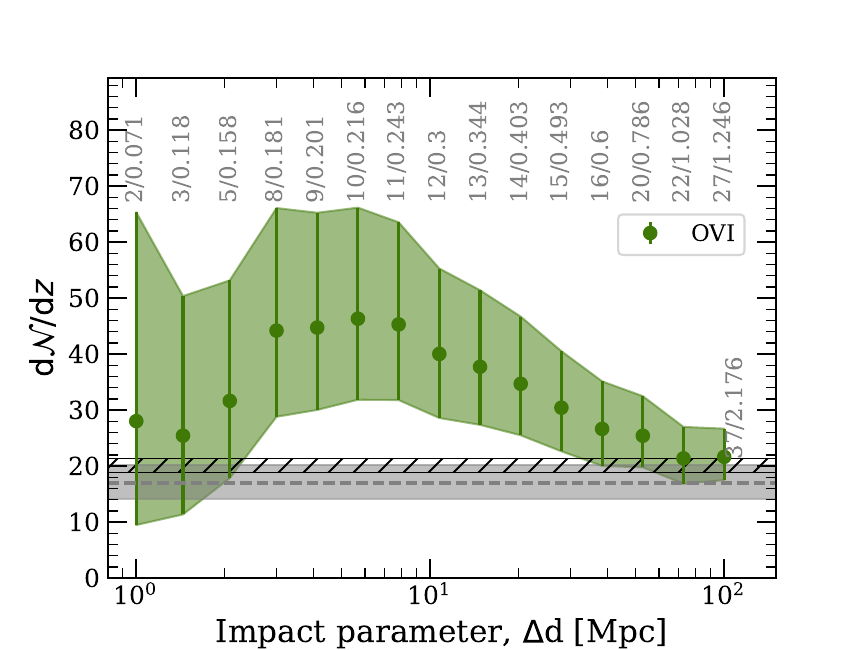}
 \end{subfigure}
    \caption{Redshift number density of our survey for total \ion{H}{I} in the top panels in \hicolor and \ion{O}{VI} in the bottom panels in \ovicolor. In the left panels, we show the redshift number density as a function of rest-frame velocity window ($\Delta v$), for a fixed $\Delta d \leq 3$ Mpc. In the right panels, we show the redshift number density as a function of the impact parameter ($\Delta d$) for a fixed rest-frame velocity window of $\Delta v = \pm 1000$ \kms. At the top of the datapoints in each panel, we indicate the total number of absorption lines detected and the redshift paths per bin in gray numbers. The expected field value estimated for our survey, $d\mathcal{N}/dz|_{field}$, is shown in the gray horizontal dashed line, and its value is indicated in the gray numbers at the end of the dashed line. The expected field values estimated from \cite{Danforth2016} are shown in black hashed regions for comparison. }
    \label{Fig:Redshift_number_density_HI_OVI}
\end{figure*}

Figure~\ref{Fig:Redshift_number_density_HI_OVI} presents the redshift number density of our survey for total \ion{H}{I} (top panels, \hicolor) and for \ion{O}{VI} (bottom panels, \ovicolor). In the left panels, we show the redshift number density as a function of the rest-frame velocity window ($\Delta v$), for a fixed $\Delta d \leq 3$ Mpc. In the right panels, we show the redshift number density as a function of the impact parameter ($\Delta d$), for a fixed velocity window of $\Delta v = \pm1000$ \kms. The ranges selected for the maximum velocity window and the maximum impact parameter were chosen to cover most of the spectrum in each sightline. The bins are not independent of each other, as indicated by the colored areas. The field redshift number density of our survey with associated uncertainties, calculated with Equation~\eqref{Eq:dNdz_field}, is shown in gray. This result is consistent with that derived from \cite{Danforth2016} ($d\mathcal{N}/dz|_{field_D}$), shown in a black hashed region. Our estimate of the expected field is $\sim125^{+9}_{-9}$, $\sim93_{-8}^{+8}$, $\sim32_{-4}^{+5}$, and $\sim17_{-3}^{+3}$ for \ion{H}{I}, NLAs, BLAs, and \ion{O}{VI}, respectively. Fixing $\Delta v=\pm1000$ \kms, we observe an overall decrease in the redshift number density of \ion{H}{I} and \ion{O}{VI} absorption lines with increasing $\Delta d$ for $\Delta d \gtrsim 3$ Mpc. In contrast, for $\Delta d \lesssim 3$ Mpc, the trend either increases or remains approximately flat with increasing $\Delta d$. This indicates the approximate location where absorbing gas is detected more frequently, therefore reinforcing the validity of our adopted fiducial value of $3$ Mpc for the maximum impact parameter. For \ion{H}{I}, with our sample of ten sightlines, we have observed an approximately flat trend for the $d\mathcal{N}/dz$ in the region $\Delta d\leq 3$ Mpc. While for \ion{O}{VI}, a slightly decreasing trend is observed with decreasing $\Delta d$ in the region $\Delta d\leq 3$ Mpc. However, we note that this behavior may be due to our small sample sizes in these bins. Fixing the impact parameter, the redshift number density generally decreases as the velocity window increases. 

\begin{figure*}
 \centering
 \begin{subfigure}[ht]{0.49\textwidth}
     \centering
     \includegraphics[width=\textwidth]{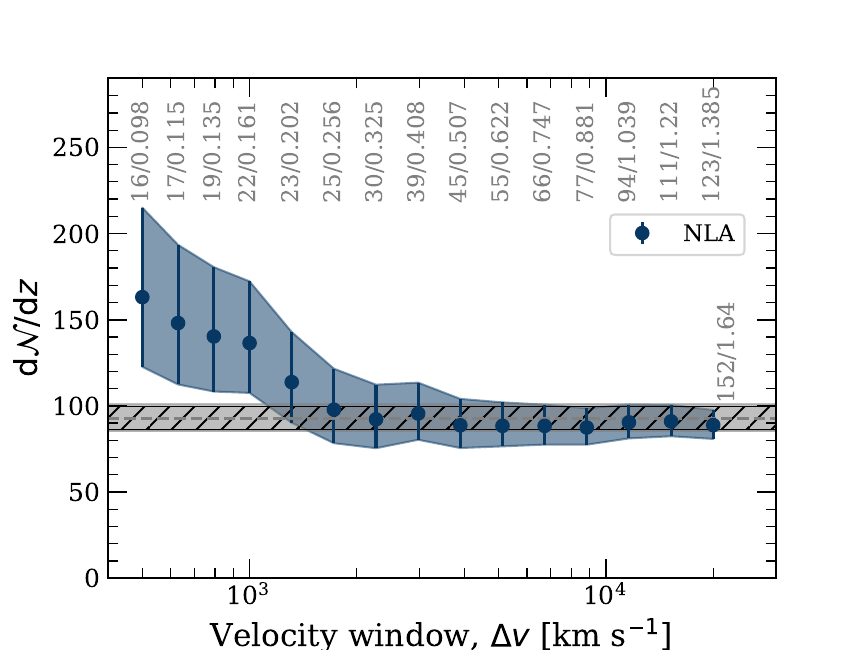}
 \end{subfigure}
 \hfill
 \begin{subfigure}[ht]{0.49\textwidth}
     \centering
     \includegraphics[width=\textwidth]{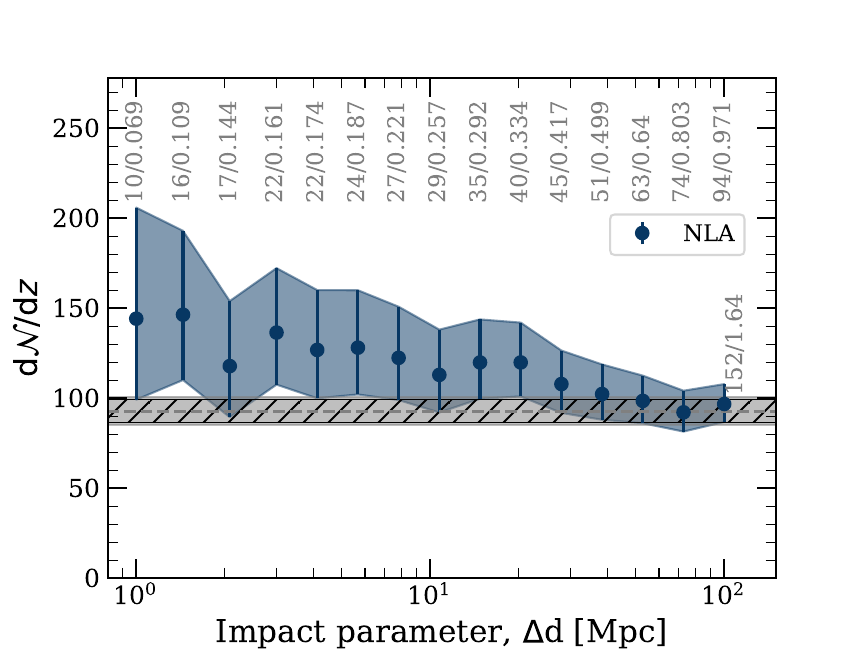}
 \end{subfigure}
 \hfill
 \begin{subfigure}[ht]{0.49\textwidth}
     \centering
     \includegraphics[width=\textwidth]{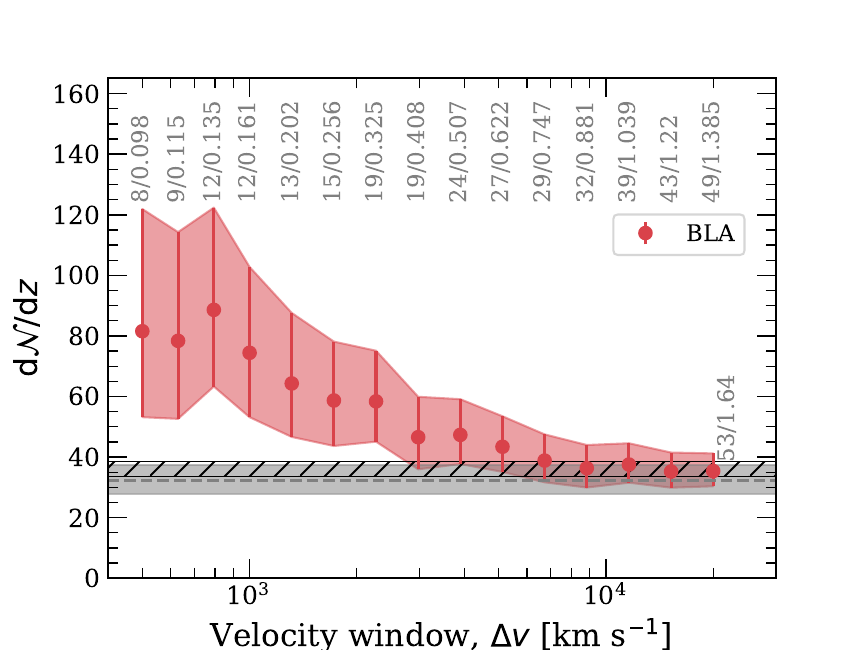}
 \end{subfigure}
 \hfill
 \begin{subfigure}[ht]{0.49\textwidth}
     \centering
     \includegraphics[width=\textwidth]{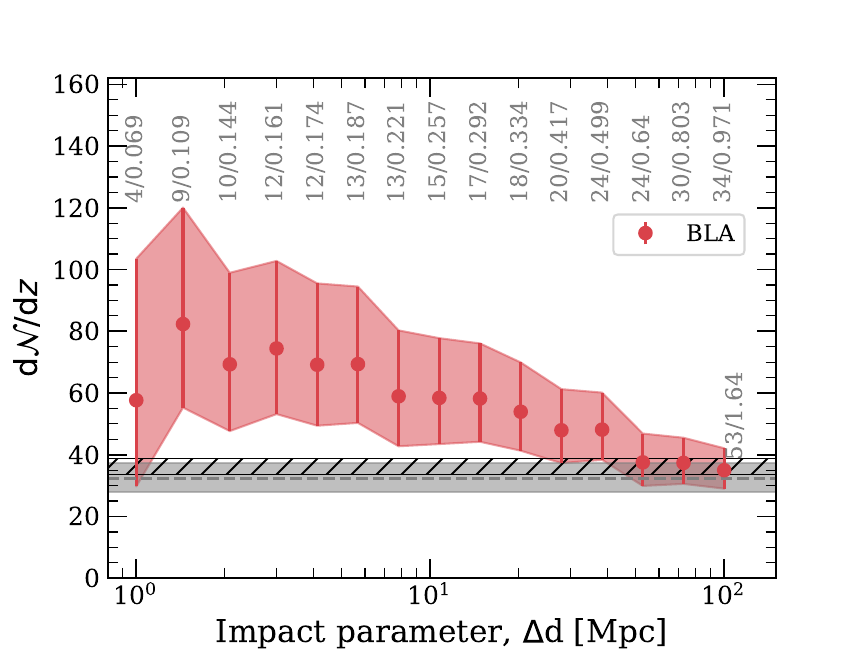}
 \end{subfigure}
    \caption{Redshift number density of our survey for NLA in the top panels in \nlacolor and BLA in the bottom panels in \blacolor. In the left panels, we show the redshift number density as a function of rest-frame velocity window ($\Delta v$), for a fixed $\Delta d \leq 3$ Mpc. In the right panels, we show the redshift number density as a function of the impact parameter ($\Delta d$) for a fixed rest-frame velocity window of $\Delta v = \pm 1000$ \kms. At the top of the datapoints in each panel, we indicate the total number of detected absorption lines and the redshift paths per bin in gray numbers. The expected field value estimated for our survey, $d\mathcal{N}/dz|_{field}$, is shown in the gray horizontal dashed line, and its value is indicated in the gray numbers at the end of the dashed line. The expected field value estimated from \cite{Danforth2016} is shown in a black hashed region for comparison.}
    \label{Fig:Redshift_number_density_NLA_BLA}
\end{figure*}

Figure~\ref{Fig:Redshift_number_density_NLA_BLA} shows the redshift number density for the samples of BLAs and NLAs as a function of the velocity window and impact parameter. In the top panels, we show the redshift number density for the NLAs sample, colored in \nlacolor. In the bottom panels, we show the redshift number density for the BLA sample, colored in \blacolor. Similar to Figure~\ref{Fig:Redshift_number_density_HI_OVI}, in the right panels, we show the redshift number density as a function of the impact parameter $\Delta d$ for a fixed velocity window $\Delta v=\pm1000$\,\kms, and in the left panels we show $d\mathcal{N}/dz$ as a function of the rest-frame velocity window $\Delta v$ for a fixed value of the impact parameter $\Delta d\leq 3$ Mpc. 
For the NLAs sample, the uncertainties are smaller than for the BLAs sample because we detected more absorbers: NLAs (152) and BLAs (53).
Finally, it is worthwhile to highlight that in Figures~\ref{Fig:Redshift_number_density_HI_OVI} and ~\ref{Fig:Redshift_number_density_NLA_BLA} the $d\mathcal{N}/dz$ approaches the expected field as both the rest-frame velocity window and impact parameter increase, as expected. 
This behavior arises because increasing the rest-frame velocity window or the impact parameter corresponds to analyzing a broader portion of the sightline, thereby reducing the restriction to regions around the targeted cluster-pairs and corresponding more to the field. 

\subsection{Relative excess of the survey with respect to the field}

Using the redshift number density of each sample of interest (total \ion{H}{I}, NLAs, BLAs, and \ion{O}{VI}), we estimate the relative excess of our absorption line samples compared to their respective field expectations. Following \cite{Tejos2016}, we defined this as $\frac{d\mathcal{N}}{dz}/\frac{d\mathcal{N}}{dz}|_{field}$ in order to compare the incidence rates of the analyzed species with their respective fields.

\begin{figure*}
 \centering
 \begin{subfigure}[ht]{0.49\textwidth}
     \centering
     \includegraphics[width=\textwidth]{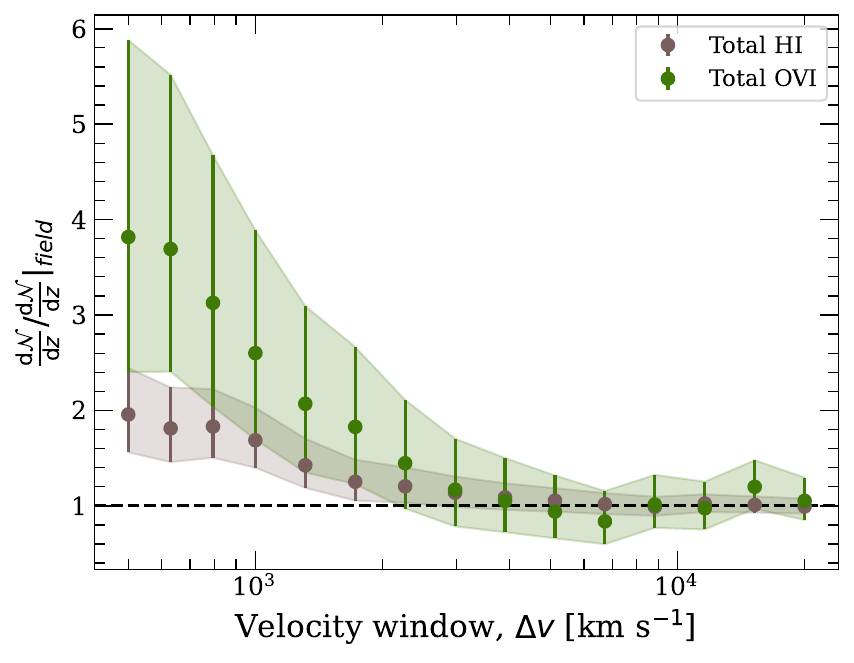}
 \end{subfigure}
 \hfill
 \begin{subfigure}[ht]{0.49\textwidth}
     \centering
     \includegraphics[width=\textwidth]{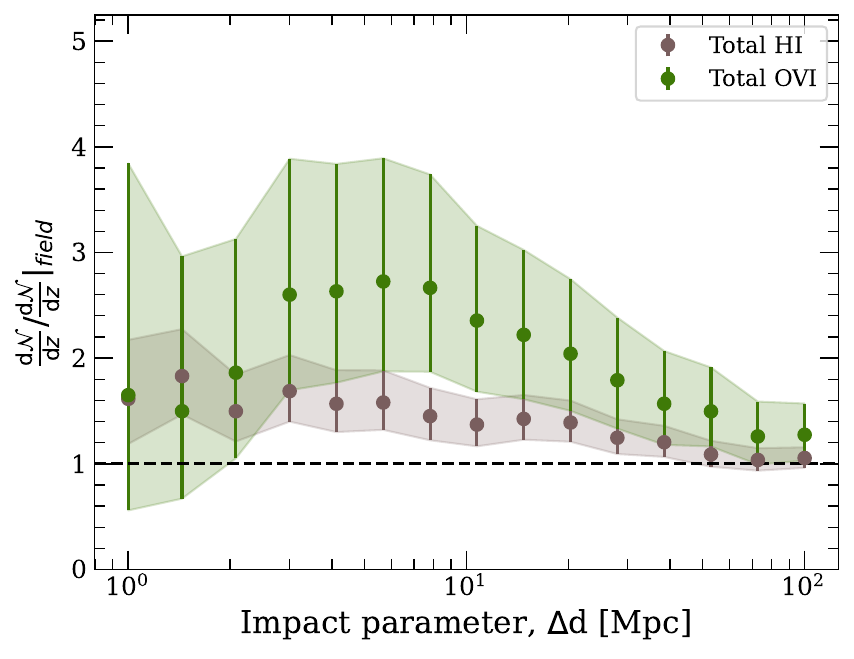}
 \end{subfigure}
 \hfill
 \begin{subfigure}[ht]{0.49\textwidth}
     \centering
     \includegraphics[width=\textwidth]{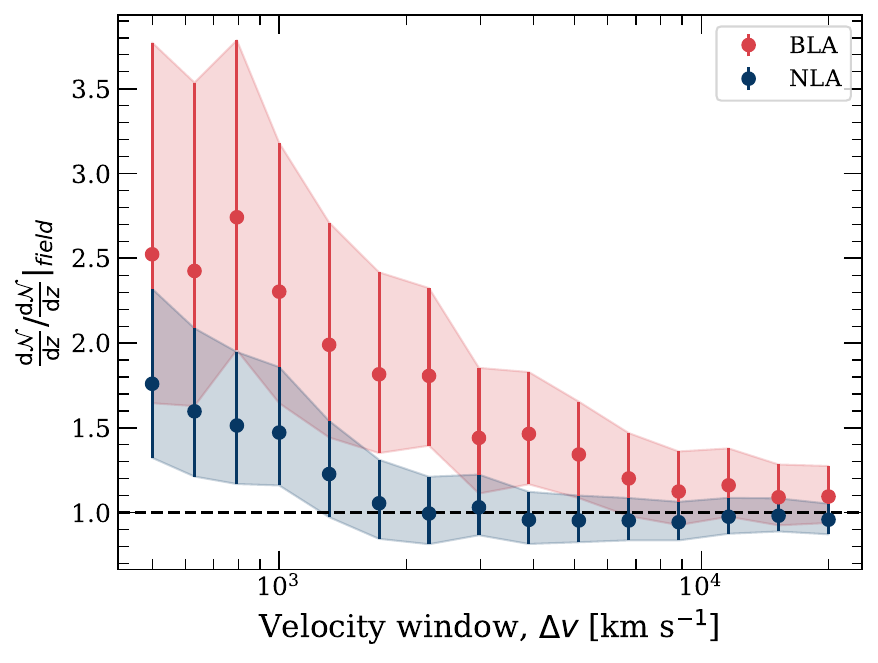}
 \end{subfigure}
 \hfill
 \begin{subfigure}[ht]{0.49\textwidth}
     \centering
     \includegraphics[width=\textwidth]{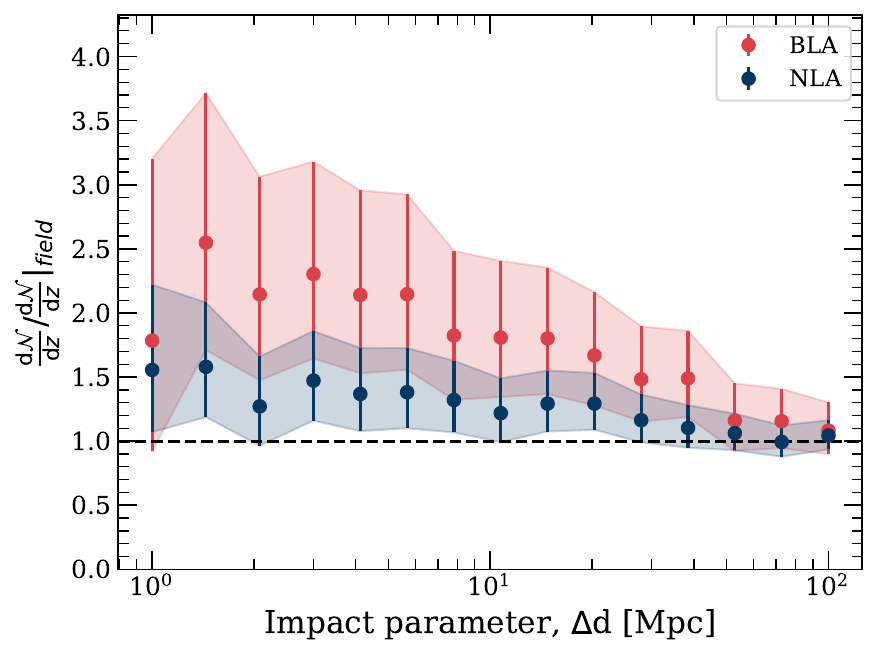}
 \end{subfigure}    
        \caption{Relative excesses of redshift number densities compared to their expected fields. They are shown as a function of rest-frame velocity window ($\Delta v$), for a fixed $\Delta d\leq3$ Mpc, in the left panels; and as a function of the impact parameter ($\Delta d$), for a fixed rest-frame velocity window $\Delta v=\pm 1000$ \kms, in the right panels. In the top panels, we show the excess for the absorption line samples of total \ion{H}{I} in \hicolor and of \ion{O}{VI} in \ovicolor. The bottom panels indicate the excess for absorption line samples of NLA in \nlacolor, and BLAs in \blacolor. Colored-shaded areas indicate the $\pm1\sigma$ statistical uncertainties.}
        \label{Fig:excess_plots}
\end{figure*}

In Figure~\ref{Fig:excess_plots}, we present the relative excess of the redshift number densities as a function of the rest-frame velocity window and the impact parameter around cluster-pairs. We indicate the excess for the total \ion{H}{I}, NLAs, BLAs, and \ion{O}{VI} with their respective $\pm1\sigma$ statistical uncertainties.  
For the fiducial values ($\Delta v=\pm1000$ \kms and $\Delta d\leq 3$ Mpc), the excess is approximately $\sim1.7_{-0.3}^{+0.3}$ for the sample of \ion{H}{I} and $\sim2.6^{+1.3}_{-0.9}$ for the sample of \ion{O}{VI}, with respect to each expected field values. The bottom panels of Figure~\ref{Fig:excess_plots} indicate that the excess of BLAs is slightly (but consistently) greater compared to the excess of NLAs at any velocity window and impact parameter. Particularly at the fiducial values, the excess of NLAs is approximately $\sim1.5^{+0.4}_{-0.3}$, and the excess of BLAs is approximately $\sim2.3^{+0.9}_{-0.7}$. The signals detected at fiducial values for total \ion{H}{I}, NLA, BLAs, and \ion{O}{VI} exhibit an excess with respect to their fields within $2.4-$, $1.4-$, $2.0-$ and $1.8-\sigma$ confidence levels, respectively.

\section{Discussion}
\label{Discussion}

\subsection{Incidence rates and relative excesses}

The redshift number density reported in this study tends to increase at smaller rest-frame velocity windows from the inter-cluster axes, as we can see in Figures~\ref{Fig:Redshift_number_density_HI_OVI} and ~\ref{Fig:Redshift_number_density_NLA_BLA}. 
However, as the rest-frame velocity window narrows, the sample size decreases, thereby increasing statistical uncertainties. Additionally, we calculated the relative excess for each species with respect to its respective field to compare the incidence rates of the analyzed species in putative filaments with those in the field. We found that the relative excess of \ion{O}{VI} is approximately twice the excess of \ion{H}{I} (Figure~\ref{Fig:excess_plots}), which is in agreement with the previous case study \cite{Tejos2016}, based on a single sightline (included in this work). For the \ion{O}{VI} sample, we obtained the largest excess compared to the field (approximately $2.6$) at $\Delta v=\pm 1000$ \kms and $\Delta d\leq 3$ Mpc; this stronger excess signal in \ion{O}{VI} may be explained by the expected higher metallicity in filaments. However, this sample has larger statistical uncertainties compared to total \ion{H}{I}, due to its small number of detections. Even though our uncertainties are large, they remain smaller than those reported by \cite{Tejos2016}. Additionally, the relative excess of the incidence rate of BLAs is systematically higher than that of NLAs. This behavior is expected if BLAs and NLAs are not probing similar physical conditions \citep{Tejos2016}. Indeed, a WHIM signature associated with inter-cluster filaments includes the detection of a relative excess of WHIM tracers (BLAs and \ion{O}{VI}) higher than that of cold gas tracers (NLAs) \citep{Tepper2012}, in line with our findings. Overall, our results on the relative excess of BLAs and \ion{O}{VI} absorbers with respect to NLAs are consistent with the predictions of cosmological hydrodynamical simulations (\citealt{Tepper2011}, \citealt{Tepper2012}), as well as with observational constraints on the density and temperature of baryons in filaments inferred from the thermal Sunyaev–Zel’dovich (tSZ) effect \citep{Erciyes2023}. They also agree with previous work by our group (\citealt{Tejos2016}, \citealt{Pessa2018}, \citealt{Pessa2025}), which showed that denser gas is preferentially located within filaments, as traced by galaxy overdensities. Finally, it is important to remind the reader that the reported incidence rates for BLAs and \ion{O}{VI} should be regarded as lower limit estimates, given the inherently conservative approach employed in this work.

\subsection{Geometric representation of our survey}

\begin{figure}
 \centering
     \includegraphics[width=0.49\textwidth]{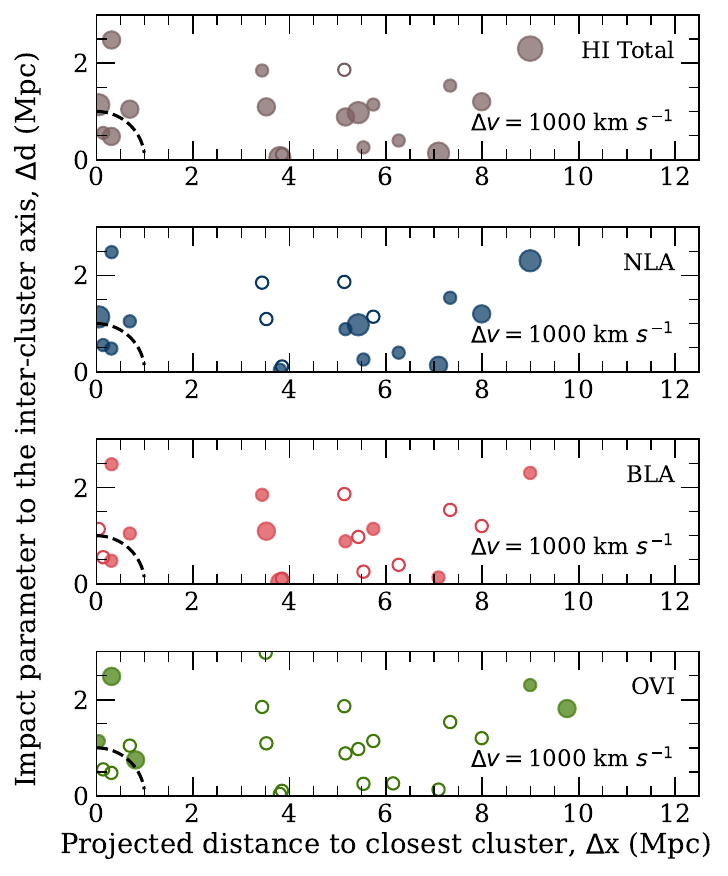}
        \caption{Representation of the position of each independent cluster-pair of our survey, relative to its corresponding QSO sightline. The y-axis indicates the impact parameter to a given cluster-pair, $\Delta d$; while the x-axis indicates the distance of the closest galaxy cluster in a given cluster-pair to the QSO sightline, known as $\Delta x$. Additionally, the approximate virial radii expected for the sample of clusters are shown as a quarter dashed circumference of radii $1$ Mpc, centered at the origin. Filled symbols correspond to sightlines showing absorption line samples of the ion species indicated in each panel within $\Delta v = \pm1000$ \kms from the redshifts of the corresponding cluster-pair. The corresponding results for $\Delta v = \pm500$ \kms are presented in Figure~\ref{Fig:geometry_appendix} in the appendix. The size of the symbols is proportional to the number of detected absorption lines. Empty symbols correspond to sightlines showing no absorbers.}
        \label{Fig:deltax_deltad}
\end{figure}

Figure~\ref{Fig:deltax_deltad} presents the location of the 22 independent cluster-pairs of our survey relative to their corresponding QSO sightline and their corresponding associated absorption features. An absorption line is associated with an independent cluster-pair if it is detected within $\Delta v$ from the cluster-pair with the smallest impact parameter of the group. This representation enables assessment of the proximity between survey sightlines and nearby clusters of the studied independent cluster-pairs, facilitating discussion of the possible origins of the observed gas excesses. On the y-axis, we indicate the impact parameter of the cluster-pair $\Delta d$, and the x-axis indicates the distance from the closest galaxy cluster in the cluster-pair to the QSO sightline along the inter-cluster axis, referred to as $\Delta x$ (see Figure~\ref{Fig:experiment}). Additionally, the approximate virial radii expected for the sample of clusters are shown as quarter dashed circumferences of radii $1$ Mpc, centered at the origin. We observe that the vast majority of sightlines are probing regions far away from the virial radii of known galaxy clusters, disfavoring the possibility that the detected gas excess is attributable to our sample of galaxy clusters. 
Filled symbols correspond to sightlines showing absorption lines of the sample indicated in each panel, within $\Delta v$ from the redshifts of a given independent cluster-pair. The size of the symbols is proportional to the number of absorption lines found. Empty symbols correspond to sightlines showing no absorbers around the cluster-pairs. 
We consider absorbers within $\Delta v = \pm1000$ \kms from the redshifts of a given independent cluster-pair. Regions where the spectra do not allow identification of a particular transition are excluded (in Figures~\ref{Fig:Survey_quality_appendix_HI} and ~\ref{Fig:Survey_quality_appendix_OVI}, we show the regions in the spectra where we expect to detect \ion{H}{I} and \ion{O}{VI} at redshifts of the independent cluster-pairs). Of the 22 independent cluster-pairs detected in our survey, we identify that 19 have \ion{H}{I} coverage, while all 22 have \ion{O}{VI} coverage. Consequently, the panels corresponding to \ion{H}{I}, NLAs, and BLAs include 19 data points, while the panel corresponding to \ion{O}{VI} includes 22 data points. In Figure~\ref{Fig:geometry_appendix} in the appendix, we show similar results but for $\Delta v = \pm500$ \kms, thus reinforcing that the signal is robust. 

\subsection{Covering fractions}

Additionally, we estimate the covering fraction of filaments close to independent cluster-pairs, $f_{c}^{filament}=n_{hits} /n_{trials}$, where $n_{hits}$ are sightlines showing absorption in a given sample, within $\Delta v =  \left\{ \pm1000, \pm500 \right\}$ \kms and within $\Delta d\leq 3$ Mpc, and $n_{trials}$ are defined as the total number of sightlines where absorbers could have been detected.

As a control sample for each absorber, we estimate the random covering fraction $f_{c}^{random}$, according to Equation~\eqref{eq:fc_random} (\citealt{Tejos2016}):
\begin{equation}
 f_{c}^{random}=\left\{\begin{matrix}
\Delta z(\Delta v) \frac{d\mathcal{N}}{dz}|_{field}&  \text{if } \Delta z(\Delta v) \frac{d\mathcal{N}}{dz}|_{field}<1\\
1 &  \text{if } \Delta z(\Delta v) \frac{d\mathcal{N}}{dz}|_{field}\geq 1\\
\end{matrix}\right.
\label{eq:fc_random}
\end{equation}

\noindent Here, $\frac{d\mathcal{N}}{dz}|_{field}$ is the expected field, $\Delta z(\Delta v)$ is the redshift path to a rest-frame velocity window $\Delta v$, evaluated at the median redshift of our independent cluster-pairs sample. For our survey, the median redshift is $0.191$ and $0.187$ for \ion{H}{I} and \ion{O}{VI}, respectively. 

To estimate the uncertainties for $f_{c}^{random}$ and $f_{c}^{filament}$, we used the Bayesian formalism assuming a confidence level of $1 \sigma$ for a binomial distribution, for the number of `hits' given the number of independent cluster-pairs `trials' for each transition \citep{Cameron2011}.

\begin{figure}
 \centering
     \includegraphics[width=0.49\textwidth]{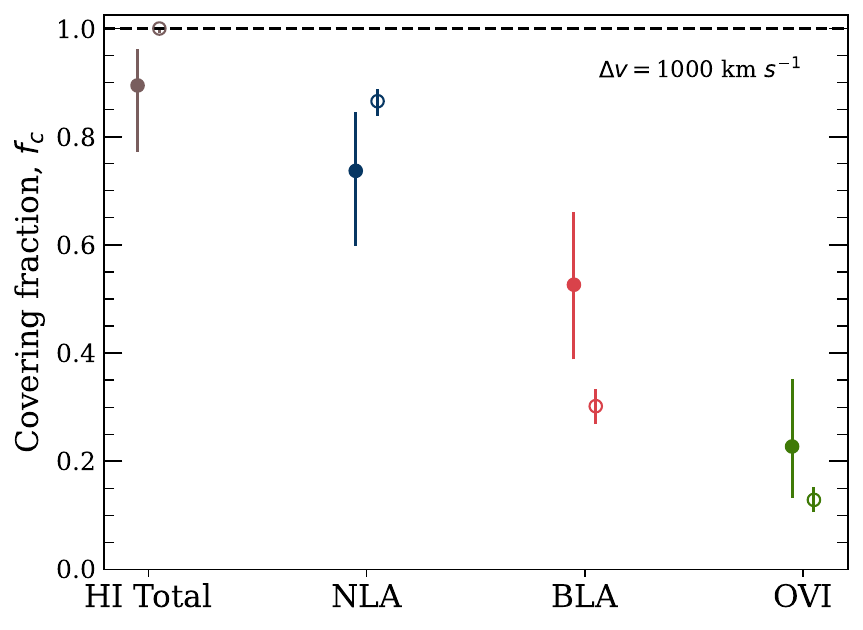}
        \caption{Covering fraction for samples of total \ion{H}{I}, NLAs, BLAs, and \ion{O}{VI} within a rest-frame velocity window of $\Delta v=\pm1000$ \kms around independent cluster-pairs. We used the same colors as in the previous plots. The $f_{c}^{filament}$ is shown in solid symbols and their corresponding randomly expected values, $f_{c}^{random}$, in open symbols. In Figure~\ref{Fig:covering_appendix} we present the covering fraction within a rest-frame velocity window of $\Delta v=\pm500$ \kms.}
        \label{Fig:covering_fraction}
\end{figure}

In Figure~\ref{Fig:covering_fraction}, we show the covering fraction of each absorber in $\Delta v=\pm1000$ \kms along the sightline close to the independent cluster-pairs named $f_{c}^{filament}$, and we compared them with the random expectation $f_{c}^{random}$ (see Table~\ref{tab:appendix_covering_fraction_1000_500} in the appendix).

The covering fraction of \ion{H}{I} at $\Delta v=\pm1000$ \kms is $f_{c}^{filament}=0.89_{-0.12}^{+0.07}$, a value close to $1$, which is in agreement with Figure~\ref{Fig:deltax_deltad}, where nearly every independent cluster-pair is filled for \ion{H}{I}.  
For the BLA and NLA samples, we obtained covering fractions of $0.53_{-0.14}^{+0.13}$ and $0.74_{-0.14}^{+0.11}$, respectively. 

For BLAs and \ion{O}{VI}, at a rest-frame velocity window of $\Delta v=\pm1000$ \kms, the covering fraction found in our study, close to cluster-pairs, is greater than their respective random expectation at the 1-$\sigma$ confidence level. However, for NLAs, the detected covering fraction is smaller than the random expectation, at the 1-$\sigma$ confidence level. For a smaller rest-frame velocity window of $\Delta v=\pm500$ \kms (see Figure~\ref{Fig:covering_appendix} in the appendix), we find that covering fractions for BLAs and \ion{O}{VI} are higher than their respective randomly expected values at the 1.7-$\sigma$ confidence level. Consequently, we have found that the excesses in the covering fraction, at both rest-frame velocity windows, for BLAs and \ion{O}{VI} are greater than the excess for NLAs. Since BLAs and \ion{O}{VI} trace temperatures of $\sim1-3\times 10^{5}$ K \citep{Richter2006b} respectively, a greater excess in these ion species than cold gas tracers (NLAs) suggests that the potential filaments associated with the independent cluster-pairs in our survey might contain the WHIM. Indeed, the increase in the covering fractions of BLAs and \ion{O}{VI} observed in this work towards inter-cluster axes, relative to random expectations, constitutes a tentative WHIM detection. 

\subsection{Potential caveats}

An important caveat when interpreting our \ion{O}{VI} results is that \ion{O}{VI} is not an unambiguous tracer of the WHIM. As discussed earlier, \ion{O}{VI} can arise in both collisionally ionized gas at temperatures characteristic of the WHIM and cooler photoionized gas with $T < 10^5$ K (\citealt{Tepper2011}, \citealt{Tripp2008}, \citealt{Danforth2008}). Therefore, although the excess of \ion{O}{VI} absorbers found in our sample is consistent with an enhanced incidence rate of warm-hot gas in the environments probed, we cannot attribute all of the \ion{O}{VI} detections to the WHIM. Notably, the simultaneous detection of BLAs, which are expected to trace warm-hot gas, provides evidence that part of the detected \ion{O}{VI} absorption may arise from the WHIM. The higher incidence rate of both \ion{O}{VI} and BLAs in the targeted environments is therefore consistent with an increased contribution from WHIM gas, although the contribution of cooler photoionized gas to the \ion{O}{VI} population cannot be ruled out

Additionally, it is possible that BLAs and NLAs could be tracing the CGM rather than the IGM because inside a filament, the chance of hitting a galaxy halo is higher than for a random sightline, which could, in principle, enhance the incidence rate relative to the field. However, galactic origin absorbers could affect BLAs and NLAs in a similar manner (\citealt{Wilde2021}, \citealt{Pessa2018}), implying that the relative excess of BLAs compared to that of NLAs would be robust. 

Furthermore, it is worth emphasizing that when characterizing absorption lines, certain biases and/or systematics can affect the analysis and/or interpretation. In addition, noise and line blends can potentially mimic broad and shallow absorption features (\citealt{Richter2006b}, \citealt{Tejos2016}, \citealt{Pessa2025}). This justifies the use of higher S/N, which enables a better characterization of the kinematic structure of the absorption features (\citealt{Richter2006b}, \citealt{Danforth2010}). This limitation is intrinsic to the absorption-line technique and, to some extent, affects any absorption-based study (see discussions in \cite{Tejos2016}). Finally, we highlight the possibility that not all BLAs and \ion{O}{VI} detected are strictly tracing the WHIM, given that \ion{O}{VI} gas may originate from individual galaxies or their immediate surroundings, and BLAs may arise from blends of multiple narrower cold components in a complex that mimics higher thermal Doppler parameters. Therefore, the possibility that a fraction of the BLA sample may correspond to unresolved or blended closely-spaced NLAs should be considered when interpreting the results.

The restrictions imposed on the catalogs were designed to increase the likelihood of detecting intergalactic structures in our survey. However, we cannot be certain that the cluster-pairs trace inter-cluster filaments because some filaments might be warped or displaced from the direct cluster–cluster axis \citep{Colberg2005}, and relevant transitions may not be detected within $\pm1000$ \kms (as in the case of QSO J1410+0910). For instance, including this QSO in the redshift number density measurement increases the redshift path value without increasing the number of detections. If we assume this QSO does not intersect a filament and we exclude it from the analysis, then we find an increase of $13\%$ in the redshift number density for both BLAs and NLAs, at the fiducial values. Therefore, the inclusion of that sightline has a minor (conservative) impact on our statistical results. Additionally, it is worth emphasizing that the trends reported in the present work remain robust even upon the exclusion of QSO J1410+2304, which was previously employed in the analysis of \cite{Tejos2016}. And even if cluster-pairs trace intergalactic filaments, low S/N may hinder detection of \ion{H}{I} and \ion{O}{VI} absorption lines. 

Finally, estimating the overdensities associated with the WHIM tracers identified in this survey is essential for assessing whether their physical properties support their association with the WHIM. A detailed characterization of these overdensities is part of ongoing work and will be presented in a forthcoming study.

\section{Summary and conclusions}
\label{conclusions}
At redshifts $z<1$, a significant fraction of baryons formerly known as `missing baryons' is predicted to reside in the warm-hot intergalactic medium (WHIM). In our experiment, we used a sample of ten QSO sightlines that meet the criteria for probing inter-cluster regions (described in Section~\ref{data}) to search for intergalactic filaments. We identified that these QSO sightlines probe 22 cluster-pairs that may define independent intergalactic structures. We analyzed the QSO sightline spectra obtained in the FUV with HST/COS and characterized those whose sightlines likely intersect these filaments. We identified all absorption lines in the spectra, not only those where inter-cluster filaments are expected to exist. We estimated observational parameters, such as the Doppler parameter and column density, via Voigt profile fitting for each transition. Using these samples, we investigated the incidence rate of two WHIM tracers: BLAs and \ion{O}{VI}. We separated the \ion{H}{I} absorbers into two samples, BLAs and NLAs, based on the width of the line, and we used a Doppler parameter $b \geq 40$ \kms for BLAs. We analyzed the redshift number density ($d\mathcal{N}/dz$), the relative excess ($\frac{d\mathcal{N}}{dz}/\frac{d\mathcal{N}}{dz}|_{field}$), and the covering fraction ($f_c^{filament}$) for total \ion{H}{I}, NLAs, BLAs, and \ion{O}{VI}.

Based on the results of the analysis of our experiment, the following conclusions are presented:
\begin{itemize}
    \item The absorption lines of WHIM tracers (BLAs and \ion{O}{VI}) identified in our survey for a velocity window $\Delta v=\pm1000$ \kms, and $\Delta d\leq3$ Mpc from the cluster-pairs are indicated in Figures~\ref{Fig:Survey_panels_BLA} and ~\ref{Fig:Survey_panels_OVI}. Eight out of the ten QSO spectra of our survey show absorption features from at least one of these two WHIM tracers. This finding suggests that we have enabled WHIM tracing using this methodology.
    \item We present the redshift number density ($d\mathcal{N}/dz$) in Figures~\ref{Fig:Redshift_number_density_HI_OVI}  and ~\ref{Fig:Redshift_number_density_NLA_BLA}. Here, we find a higher incidence rate than expected in the field near independent cluster-pairs at velocity windows $\Delta v=\pm1000$ \kms, suggesting the presence of intergalactic filaments between cluster-pairs. From Figure~\ref{Fig:excess_plots}, we conclude that all the tracers have an excess absorption due to putative filaments. Indeed, the excess of \ion{O}{VI} is around three times the expected field value within approximately a $1.8-\sigma$ confidence level; and for BLAs, the excess is around twice the expected field value within approximately a $2.0-\sigma$ confidence level. These values should be regarded as lower limit estimates, given the inherently conservative approach employed throughout this study. We argued that the larger relative excess detected for BLAs and \ion{O}{VI} in regions near cluster-pairs, compared with NLAs, may be an indirect signature of the WHIM. Additionally, we highlight that the reported excesses may be larger given potential impurities in the cluster-pair sample and excess densities in the field sample.
    \item From Figure~\ref{Fig:covering_fraction}, we also conclude that for WHIM tracers (BLAs and \ion{O}{VI}) there appears to be a higher covering fraction close to cluster-pairs compared to a random control sample to 1-$\sigma$ confidence. At a rest-frame velocity window of $\Delta v=\pm1000$ \kms and $\Delta v=\pm500$ \kms (see Figure~\ref{Fig:covering_appendix} in the appendix), the excesses in the covering fraction of BLAs and \ion{O}{VI} are larger than those of NLAs. This implies that near the inter-cluster axis, there is a higher likelihood of detecting the WHIM using this technique. It is also noteworthy that the covering fraction of NLAs drops more when going from $\Delta v=\pm1000$ \kms to $\Delta v=\pm500$ \kms than that of BLAs and \ion{O}{VI}. This further indicates that the BLAs and \ion{O}{VI} are more strongly correlated with the filaments.
\end{itemize}

Based on ten sightlines, this work corroborates the trends previously reported by \cite{Tejos2016} from a single sightline, demonstrating the robustness of these findings, and the reported incidence rates reaffirm a tentative excess of WHIM tracers near inter-cluster filaments. As future work, we expect to present a second article that includes a study of the overdensities observed in this survey. Finally, we aim to further explore our results by reproducing the experiment with cosmological simulations of the low-redshift intergalactic medium, not previously studied in this context.

\section*{Data availability}
The complete version of Table~\ref{tab: parametros} is only available in electronic form at CDS.

\begin{acknowledgements}
      We sincerely thank the anonymous referee for their valuable comments and constructive suggestions, which enable a deeper discussion in the paper. 
      This research is based on data obtained from the NASA/ESA Hubble Space Telescope under programs GO~15293 and GO~12958, obtained at the Space Telescope Science Institute and from the Mikulski Archive for Space Telescopes (MAST). STScI is operated by the Association of Universities for Research in Astronomy, Inc., under NASA contract NAS5-26555. 
      K.M.A. acknowledges support from the National Agency for Research and Development (ANID)/Scholarship Program/Doctorado Nacional/2022-21220649 and `Beca Término de Tesis 2025' from PUCV. N.T. acknowledges support from FONDECYT grant 1252229. I.P. acknowledges funding by the European Research Council through ERC-AdG SPECMAP-CGM, GA~101020943. N.T., J.X.P. and J.W. acknowledge support from NASA program HST-GO-15293. S. L. acknowledges support from FONDECYT grant 1231187. We thank R. Bielby, R. Gonzalez, C. Hummels, S. L. Morris and T. Theuns for their contribution to the early stages of this project. We thank J. C. Howk for useful discussions.
\end{acknowledgements}

\bibliographystyle{aa}
\bibliography{bibliography}

\begin{appendix}
\nolinenumbers
\section{Quality of our survey}
\label{sec:appendix_survey_quality}

\begin{figure*}
\centering
\includegraphics[width=0.87\textwidth]{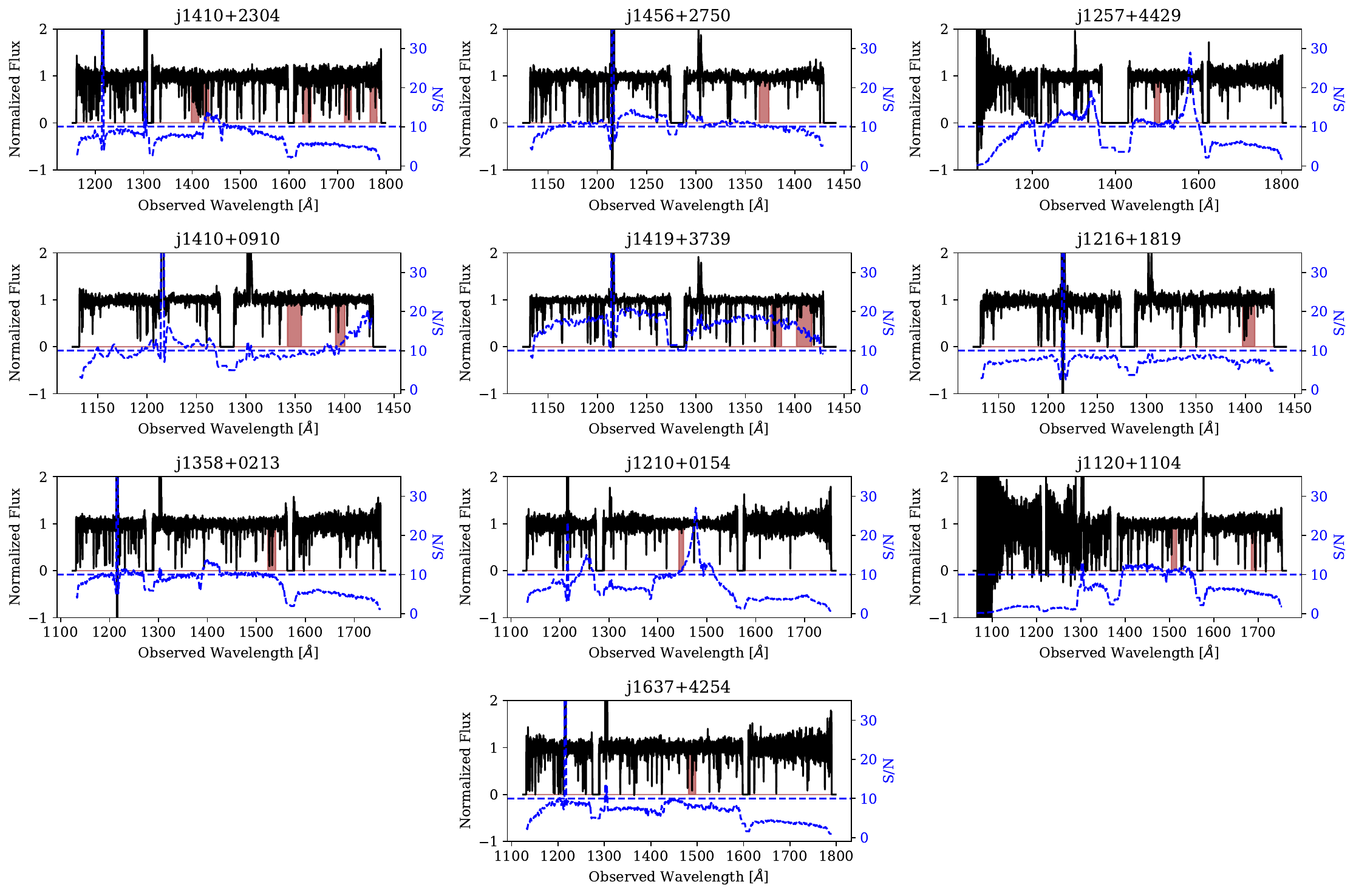}
  \caption{Quality of our survey for the \ion{H}{I} sample. The S/N per resolution element is shown in blue (values shown on the rightward y-axis) to indicate the quality of the spectra (shown in black) of our survey for \ion{H}{I}. The straight blue-dashed line corresponds to an S/N $=10$. The regions where the independent cluster-pairs are located in the spectra for the \ion{H}{I} are shown in \hicolor.}
  \label{Fig:Survey_quality_appendix_HI}
\end{figure*}

\begin{figure*}
\centering
\includegraphics[width=0.87\textwidth]{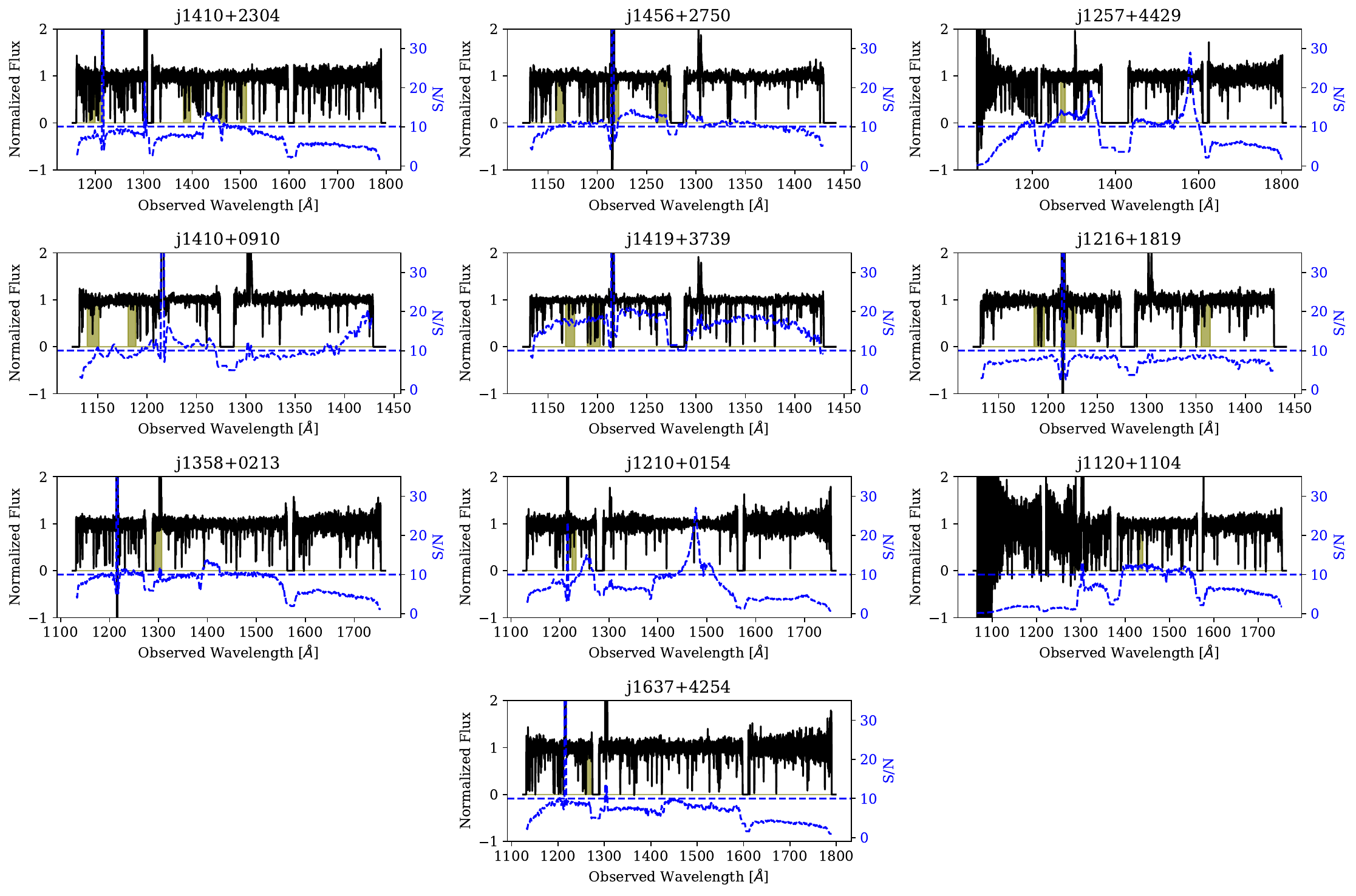}
  \caption{Quality of our survey for the \ion{O}{VI} sample. The S/N per resolution element is shown in blue (values shown on the rightward y-axis) to indicate the quality of the spectra (shown in black) of our survey for \ion{O}{VI}. The straight blue-dashed line corresponds to an S/N $=10$. The regions where the independent cluster-pairs are located in the spectra for the \ion{O}{VI} are shown in \ovicolor.}
  \label{Fig:Survey_quality_appendix_OVI}
\end{figure*}

Figures ~\ref{Fig:Survey_quality_appendix_HI} and ~\ref{Fig:Survey_quality_appendix_OVI} show the S/N of the spectra of the QSO sightlines of our survey. In each sightline, the colored regions belong to the expected observed wavelength for \ion{H}{I} (Figure~\ref{Fig:Survey_quality_appendix_HI}, in \hicolor) and \ion{O}{VI} (Figure~\ref{Fig:Survey_quality_appendix_OVI}, in \ovicolor) detections, at the location of independent cluster-pairs. 

\section{Table cluster-pairs}
\label{sec:appendix_cluster-pairs}

In Table~\ref{tab:cl-pairs}, we show the galaxy cluster-pairs associated with their respective QSO sightlines and their characteristics mentioned in section~\ref{data}.

\begin{table*} 
\caption{Properties of the cluster-pairs associated with the QSO sightlines}
\label{tab:cl-pairs}
\tiny
 \centering 
 \begin{tabular}{cccccccc} 
 \hline\hline
Pair ID&QSO&$z_{\text{pair}}$&Transverse (Mpc)&LOS (\kms)&$\Delta d$ (Mpc)&$\Delta x$ (Mpc)&Grouped ID\\ 
(1)&(2)&(3)&(4)&(5)&(6)&(7)&(8)\\ \hline 
1&J1410+2304&0.15426&22.1&1952&0.6&0.1&1\\ 
2&J1410+2304&0.15883&14.7&420&0.5&0.3&2\\ 
3&J1410+2304&0.17598&14.7&1381&1.5&7.3&3\\ 
4&J1410+2304&0.34387&12.7&1192&1.8&3.4&4\\ 
5&J1410+2304&0.34884&18.1&1022&2.3&9.0&4\\ 
6&J1410+2304&0.41487&5.3&435&1.9&2.0&5\\ 
7&J1410+2304&0.41631&16.7&1044&1.2&3.6&5\\ 
8&J1410+2304&0.41688&23.6&1287&0.0&3.8&5\\ 
9&J1410+2304&0.45838&11.1&68&1.0&0.7&6\\ 
10&J1410+2304&0.45988&16.2&682&1.2&0.2&6\\ 
11&J1410+2304&0.46019&21.9&807&1.0&0.7&6\\ 
12&J1456+2750&0.12588&17.6&207&0.9&5.2&7\\ 
13&J1456+2750&0.12628&10.2&5&1.6&5.0&7\\ 
14&J1456+2750&0.17993&15.0&312&3.0&3.5&8\\ 
15&J1456+2750&0.22637&7.5&24&0.8&0.8&9\\ 
16&J1456+2750&0.22656&15.7&116&0.8&0.8&9\\ 
17&J1456+2750&0.22687&9.1&267&1.0&0.4&9\\ 
18&J1257+4429&0.23287&17.2&32&2.3&8.1&10\\ 
19&J1257+4429&0.23306&11.6&123&1.7&3.3&10\\ 
20&J1257+4429&0.23518&21.6&1087&2.9&8.9&10\\ 
21&J1257+4429&0.23526&22.2&947&2.4&2.8&10\\ 
22&J1257+4429&0.23536&16.5&996&1.1&3.5&10\\ 
23&J1410+0910&0.10781&13.2&390&1.4&3.6&11\\ 
24&J1410+0910&0.10817&18.7&196&1.6&3.5&11\\ 
25&J1410+0910&0.10868&23.2&84&2.0&3.3&11\\ 
26&J1410+0910&0.10889&22.9&197&2.0&3.3&11\\ 
27&J1410+0910&0.11097&13.7&1315&0.5&3.8&11\\ 
28&J1410+0910&0.11178&16.1&1751&0.1&3.8&11\\ 
29&J1410+0910&0.11194&22.9&791&2.6&9.7&11\\ 
30&J1410+0910&0.14801&25.0&1341&1.9&5.1&12\\ 
31&J1419+3739&0.13551&1.8&323&2.5&0.3&13\\ 
32&J1419+3739&0.13683&16.0&1017&2.7&1.1&13\\ 
33&J1419+3739&0.15687&11.2&766&1.0&5.4&14\\ 
34&J1419+3739&0.15889&17.1&1812&2.9&4.7&14\\ 
35&J1419+3739&0.15985&13.9&775&2.3&5.4&14\\ 
36&J1419+3739&0.16187&21.1&270&2.4&8.5&14\\ 
37&J1216+1819&0.15306&22.0&1342&1.2&8.0&15\\ 
38&J1216+1819&0.15556&11.2&39&1.6&3.3&15\\ 
39&J1216+1819&0.18143&6.2&798&2.3&0.5&16\\ 
40&J1216+1819&0.18237&21.9&1274&0.3&6.2&16\\ 
41&J1216+1819&0.18394&16.2&477&2.3&0.6&16\\ 
42&J1216+1819&0.1866&6.7&1820&0.3&2.3&16\\ 
43&J1216+1819&0.31771&24.7&1455&1.8&9.8&17\\ 
44&J1358+0213&0.25746&13.2&473&1.7&5.7&18\\ 
45&J1358+0213&0.26055&20.9&481&2.0&5.5&18\\ 
46&J1358+0213&0.2612&11.0&169&1.4&5.4&18\\ 
47&J1358+0213&0.26136&20.8&246&0.3&5.5&18\\ 
48&J1358+0213&0.26209&16.8&255&1.5&5.6&18\\ 
49&J1210+0154&0.19083&13.0&1242&1.1&5.7&19\\ 
50&J1120+1104&0.24151&20.7&484&2.8&7.5&20\\ 
51&J1120+1104&0.24219&14.3&814&0.4&6.3&20\\ 
52&J1120+1104&0.24322&19.2&317&1.1&6.2&20\\ 
53&J1120+1104&0.39072&17.4&1419&1.1&0.0&21\\ 
54&J1637+4254&0.22277&15.5&512&1.9&4.8&22\\ 
55&J1637+4254&0.22366&15.2&773&2.5&6.6&22\\ 
56&J1637+4254&0.22372&18.6&802&1.2&8.8&22\\ 
57&J1637+4254&0.22453&12.2&349&0.6&5.1&22\\ 
58&J1637+4254&0.22459&14.9&378&0.7&5.1&22\\ 
59&J1637+4254&0.22521&19.0&1705&1.5&8.1&22\\ 
60&J1637+4254&0.22697&15.4&845&0.1&7.1&22\\ 
61&J1637+4254&0.22703&17.7&816&1.9&8.0&22\\ 
\hline 
 \end{tabular} 
 \tablefoot{(1) Pair ID. (2) QSO associated with a cluster-pair. (3) Redshift of the cluster-pair. (4) Transverse separation between cluster-pairs. (5) Separation between the cluster-pairs in rest-frame velocity along the QSO sightline (LOS). (6) Impact parameter ($\Delta d$). (7) Distance to the closest cluster of the pair, along the inter-cluster axis ($\Delta x$). (8) Grouped independent cluster-pairs.} 
\end{table*}

\section{Table absorption features}
\label{sec:appendix_absorptions}

In Table~\ref{tab: parametros}, we present the observational parameters derived from a subset of the absorption lines detected and fitted in all QSO sightlines of our survey (see section~\ref{analysis}). The complete table is available at the CDS.

\begin{table*} 
\caption{Observational parameters of the absorption features detected in our survey}
\label{tab: parametros}
 \tiny 
 \centering 
 \begin{tabular}{cccccccccc} 
 \hline \hline
ID&ion&$\lambda_{obs}/$\AA&z&log($N/cm^{-2}$)&$b$/\kms&$W_r/$\AA&S/N&reliability&QSO\\ 
$(1)$&$(2)$&$(3)$&$(4)$&$(5)$&$(6)$&$(7)$&$(8)$&$(9)$&$(10)$\\  \hline 
$1$&N V&$1238.6$&$-0.00020\pm0.00003$&$13.62\pm0.09$&$41\pm12$&$0.080\pm0.017$&$13\pm1$&a&J1410+2304\\ 
$2$&Si IV&$1393.5$&$-0.00020\pm0.00001$&$13.78\pm0.03$&$39\pm3$&$0.334\pm0.024$&$11\pm1$&a&J1410+2304\\ 
$3$&Si III&$1206.3$&$-0.00018\pm0.00001$&$13.87\pm0.13$&$37\pm5$&$0.437\pm0.081$&$11\pm1$&b&J1410+2304\\ 
$4$&C IV&$1548.0$&$-0.00016\pm0.00001$&$14.28\pm0.03$&$38\pm3$&$0.419\pm0.030$&$12\pm1$&a&J1410+2304\\ 
$5$&Si II&$1260.2$&$-0.00015\pm0.00001$&$14.57\pm0.03$&$37\pm1$&$0.581\pm0.020$&$14\pm1$&a&J1410+2304\\ 
$6$&Al II&$1670.6$&$-0.00014\pm0.00001$&$13.35\pm0.08$&$32\pm4$&$0.434\pm0.062$&$8\pm1$&b&J1410+2304\\ 
$7$&S II&$1259.4$&$-0.00013\pm0.00001$&$15.16\pm0.04$&$31\pm3$&$0.224\pm0.020$&$14\pm1$&a&J1410+2304\\ 
$8$&C II&$1334.4$&$-0.00013\pm0.00001$&$15.13\pm0.12$&$47\pm4$&$0.644\pm0.086$&$10\pm1$&b&J1410+2304\\ 
$9$&C II*&$1335.6$&$-0.00009\pm0.00002$&$13.99\pm0.07$&$31\pm8$&$0.142\pm0.025$&$10\pm1$&b&J1410+2304\\ 
$10$&Si III&$1206.7$&$0.00013\pm0.00001$&$13.04\pm0.07$&$24\pm6$&$0.157\pm0.031$&$11\pm1$&b&J1410+2304\\ 
$11$&N I&$1200.0$&$0.00040\pm0.00001$&$13.76\pm0.06$&$23\pm6$&$0.080\pm0.014$&$13\pm2$&a&J1410+2304\\ 
$12$&Si III&$1207.1$&$0.00050\pm0.00003$&$12.49\pm0.13$&$27\pm14$&$0.058\pm0.019$&$11\pm1$&b&J1410+2304\\ 
$13$&H I&$1218.0$&$0.00194\pm0.00003$&$13.11\pm0.21$&$15\pm16$&$0.058\pm0.045$&$8\pm2$&c&J1410+2304\\ 
$14$&H I&$1224.7$&$0.00740\pm0.00002$&$12.99\pm0.13$&$15\pm11$&$0.046\pm0.017$&$12\pm1$&c&J1410+2304\\ 
$15$&H I&$1249.0$&$0.02745\pm0.00003$&$13.52\pm0.12$&$35\pm11$&$0.144\pm0.040$&$14\pm1$&b&J1410+2304\\ 
$16$&H I&$1250.5$&$0.02863\pm0.00002$&$13.12\pm0.15$&$10\pm9$&$0.054\pm0.030$&$14\pm1$&c&J1410+2304\\ 
$17$&H I&$1250.8$&$0.02890\pm0.00001$&$13.58\pm0.05$&$35\pm6$&$0.160\pm0.020$&$14\pm1$&b&J1410+2304\\ 
$18$&H I&$1254.8$&$0.03222\pm0.00006$&$13.03\pm0.18$&$49\pm28$&$0.055\pm0.023$&$14\pm1$&c&J1410+2304\\ 
$19$&H I&$1255.9$&$0.03307\pm0.00004$&$13.46\pm0.10$&$55\pm12$&$0.139\pm0.030$&$14\pm1$&b&J1410+2304\\ 
$20$&H I&$1261.0$&$0.03729\pm0.00002$&$12.82\pm0.17$&$7\pm12$&$0.029\pm0.040$&$14\pm1$&c&J1410+2304\\ 
$21$&H I&$1262.2$&$0.03830\pm0.00002$&$12.95\pm0.12$&$12\pm9$&$0.041\pm0.015$&$14\pm1$&c&J1410+2304\\ 
$22$&H I&$1264.2$&$0.03991\pm0.00003$&$12.74\pm0.19$&$17\pm16$&$0.028\pm0.016$&$13\pm1$&c&J1410+2304\\ 
$23$&H I&$1266.8$&$0.04206\pm0.00003$&$13.47\pm0.10$&$35\pm10$&$0.132\pm0.031$&$13\pm1$&b&J1410+2304\\ 
$24$&H I&$1268.8$&$0.04374\pm0.00001$&$13.12\pm0.18$&$8\pm7$&$0.050\pm0.034$&$13\pm1$&c&J1410+2304\\ 
$25$&H I&$1269.3$&$0.04415\pm0.00004$&$13.01\pm0.14$&$36\pm17$&$0.052\pm0.017$&$13\pm1$&b&J1410+2304\\ 
$26$&H I&$1292.0$&$0.06279\pm0.00001$&$13.34\pm0.07$&$22\pm6$&$0.094\pm0.018$&$12\pm1$&b&J1410+2304\\ 
$27$&H I&$1322.2$&$0.08763\pm0.00003$&$15.56\pm1.99$&$1\pm6$&$0.052\pm0.097$&$8\pm1$&c&J1410+2304\\ 
$28$&H I&$1324.8$&$0.08976\pm0.00003$&$13.13\pm0.14$&$25\pm13$&$0.064\pm0.023$&$9\pm1$&c&J1410+2304\\ 
$29$&H I&$1335.2$&$0.09832\pm0.00002$&$13.11\pm0.12$&$14\pm8$&$0.056\pm0.020$&$10\pm1$&c&J1410+2304\\ 
$30$&H I&$1336.1$&$0.09910\pm0.00003$&$13.20\pm0.12$&$34\pm14$&$0.077\pm0.022$&$10\pm1$&b&J1410+2304\\ 
$31$&H I&$1336.6$&$0.09947\pm0.00002$&$13.02\pm0.14$&$14\pm9$&$0.047\pm0.018$&$10\pm1$&c&J1410+2304\\ 
$32$&H I&$1370.0$&$0.12697\pm0.00003$&$12.92\pm0.18$&$21\pm14$&$0.041\pm0.018$&$11\pm1$&c&J1410+2304\\ 
$33$&H I&$1378.5$&$0.13396\pm0.00001$&$12.96\pm0.21$&$22\pm14$&$0.044\pm0.022$&$12\pm1$&c&J1410+2304\\ 
$34$&H I&$1380.5$&$0.13560\pm0.00005$&$13.19\pm0.12$&$51\pm18$&$0.078\pm0.021$&$12\pm1$&b&J1410+2304\\ 
$35$&H I&$1392.6$&$0.14556\pm0.00001$&$13.08\pm0.11$&$13\pm7$&$0.052\pm0.016$&$11\pm1$&b&J1410+2304\\ 
$36$&H I&$1400.4$&$0.15193\pm0.00006$&$13.12\pm0.16$&$48\pm24$&$0.067\pm0.025$&$11\pm1$&c&J1410+2304\\ 
$37$&H I&$1401.0$&$0.15243\pm0.00001$&$13.54\pm0.06$&$29\pm6$&$0.145\pm0.021$&$11\pm1$&b&J1410+2304\\ 
$38$&H I&$1410.9$&$0.16061\pm0.00001$&$13.32\pm0.11$&$67\pm22$&$0.105\pm0.027$&$13\pm1$&b&J1410+2304\\ 
$39$&H I&$1412.6$&$0.16199\pm0.00002$&$12.95\pm0.13$&$16\pm9$&$0.043\pm0.014$&$13\pm1$&b&J1410+2304\\ 
$40$&H I&$1419.1$&$0.16737\pm0.00010$&$13.03\pm0.19$&$66\pm36$&$0.056\pm0.025$&$12\pm1$&c&J1410+2304\\ 
$41$&H I&$1420.3$&$0.16831\pm0.00007$&$12.48\pm0.45$&$16\pm30$&$0.016\pm0.019$&$11\pm1$&c&J1410+2304\\ 
$42$&H I&$1420.6$&$0.16859\pm0.00001$&$13.22\pm0.14$&$31\pm14$&$0.078\pm0.027$&$11\pm1$&c&J1410+2304\\ 
$43$&H I&$1424.6$&$0.17188\pm0.00004$&$12.82\pm0.15$&$26\pm14$&$0.034\pm0.013$&$16\pm1$&c&J1410+2304\\ 
$44$&H I&$1433.2$&$0.17890\pm0.00001$&$13.01\pm0.08$&$13\pm5$&$0.046\pm0.009$&$18\pm2$&b&J1410+2304\\ 
$45$&H I&$1440.0$&$0.18453\pm0.00010$&$13.20\pm0.17$&$70\pm29$&$0.082\pm0.032$&$18\pm1$&c&J1410+2304\\ 
$46$&Si III&$1429.3$&$0.18466\pm0.00003$&$12.42\pm0.10$&$38\pm12$&$0.052\pm0.012$&$19\pm1$&b&J1410+2304\\ 
$47$&H I&$1441.0$&$0.18539\pm0.00002$&$13.06\pm0.08$&$28\pm7$&$0.056\pm0.010$&$18\pm1$&b&J1410+2304\\ 
$48$&H I&$1444.1$&$0.18791\pm0.00001$&$13.44\pm0.07$&$106\pm20$&$0.139\pm0.022$&$18\pm1$&b&J1410+2304\\ 
$49$&H I&$1449.9$&$0.19269\pm0.00004$&$12.83\pm0.14$&$35\pm17$&$0.035\pm0.012$&$17\pm1$&c&J1410+2304\\ 
$50$&H I&$1460.0$&$0.20099\pm0.00005$&$12.82\pm0.24$&$24\pm21$&$0.034\pm0.021$&$10\pm2$&c&J1410+2304\\ 
$51$&H I&$1469.8$&$0.20904\pm0.00001$&$15.45\pm0.20$&$15\pm1$&$0.276\pm0.033$&$15\pm1$&a&J1410+2304\\ 
$52$&C II&$1614.1$&$0.20951\pm0.00001$&$14.38\pm0.08$&$17\pm4$&$0.197\pm0.039$&$5\pm1$&a&J1410+2304\\ 
$53$&Si IV&$1685.8$&$0.20955\pm0.00001$&$13.63\pm0.06$&$17\pm3$&$0.186\pm0.031$&$8\pm1$&a&J1410+2304\\ 
$54$&N III&$1197.2$&$0.20955\pm0.00001$&$14.40\pm0.12$&$17\pm5$&$0.133\pm0.039$&$11\pm1$&b&J1410+2304\\ 
$55$&Si II&$1524.5$&$0.20955\pm0.00001$&$12.97\pm0.10$&$8\pm3$&$0.078\pm0.026$&$12\pm1$&a&J1410+2304\\ 
$56$&N V&$1498.5$&$0.20959\pm0.00002$&$13.59\pm0.11$&$25\pm8$&$0.072\pm0.019$&$15\pm1$&a&J1410+2304\\ 
$57$&Si III&$1459.4$&$0.20961\pm0.00001$&$13.56\pm0.05$&$32\pm3$&$0.323\pm0.035$&$11\pm2$&b&J1410+2304\\ 
$58$&O VI&$1248.2$&$0.20962\pm0.00005$&$14.18\pm0.76$&$27\pm17$&$0.135\pm0.155$&$14\pm1$&a&J1410+2304\\ 
$59$&C III&$1181.8$&$0.20963\pm0.00003$&$16.77\pm0.42$&$17\pm3$&$0.422\pm0.161$&$11\pm1$&c&J1410+2304\\ 
$60$&H I&$1470.6$&$0.20966\pm0.00001$&$17.53\pm0.23$&$21\pm2$&$0.688\pm0.112$&$15\pm1$&a&J1410+2304\\ 
$61$&Si II&$1524.7$&$0.20968\pm0.00001$&$13.02\pm0.10$&$42\pm8$&$0.145\pm0.032$&$12\pm1$&a&J1410+2304\\ 
$62$&C II&$1614.4$&$0.20975\pm0.00001$&$14.07\pm0.10$&$27\pm8$&$0.171\pm0.042$&$5\pm1$&a&J1410+2304\\ 
$63$&N III&$1197.5$&$0.20980\pm0.00003$&$13.99\pm0.12$&$25\pm12$&$0.083\pm0.026$&$11\pm1$&b&J1410+2304\\ 
$64$&Si IV&$1686.2$&$0.20981\pm0.00002$&$13.22\pm0.08$&$23\pm7$&$0.115\pm0.025$&$8\pm1$&a&J1410+2304\\ 
$65$&O VI&$1248.5$&$0.20983\pm0.00017$&$14.39\pm0.52$&$49\pm46$&$0.224\pm0.237$&$14\pm1$&a&J1410+2304\\ 
$66$&Si III&$1459.7$&$0.20985\pm0.00001$&$13.04\pm0.10$&$12\pm4$&$0.113\pm0.036$&$11\pm2$&b&J1410+2304\\ 
$67$&C III&$1182.1$&$0.20986\pm0.00032$&$13.61\pm1.19$&$36\pm54$&$0.181\pm0.293$&$10\pm1$&c&J1410+2304\\ 
$68$&O VI&$1248.8$&$0.21018\pm0.00001$&$14.10\pm0.15$&$17\pm7$&$0.105\pm0.039$&$14\pm1$&a&J1410+2304\\ 
$69$&N V&$1499.2$&$0.21022\pm0.00004$&$13.67\pm0.09$&$60\pm19$&$0.093\pm0.021$&$14\pm1$&a&J1410+2304\\  
$continue$ $online...$& & & & & & &\\
\hline 
 \end{tabular} 
\tablefoot{For each identification, we report the (1) ID, (2) name of the ion, (3) wavelength of observation, (4) redshift, (5) column density, (6) Doppler parameter, (7) equivalent width, (8) signal-to-noise ratio, and (9) reliability. We present only the first few rows, corresponding to absorption features in QSO J1210+0154 sightline; the complete table is available in a digital repository. } 
 \end{table*}

\section{Oxygen transitions}
\label{sec:appendix_OVI_transition}

\begin{figure*}
\includegraphics[width=\textwidth,height=12cm]{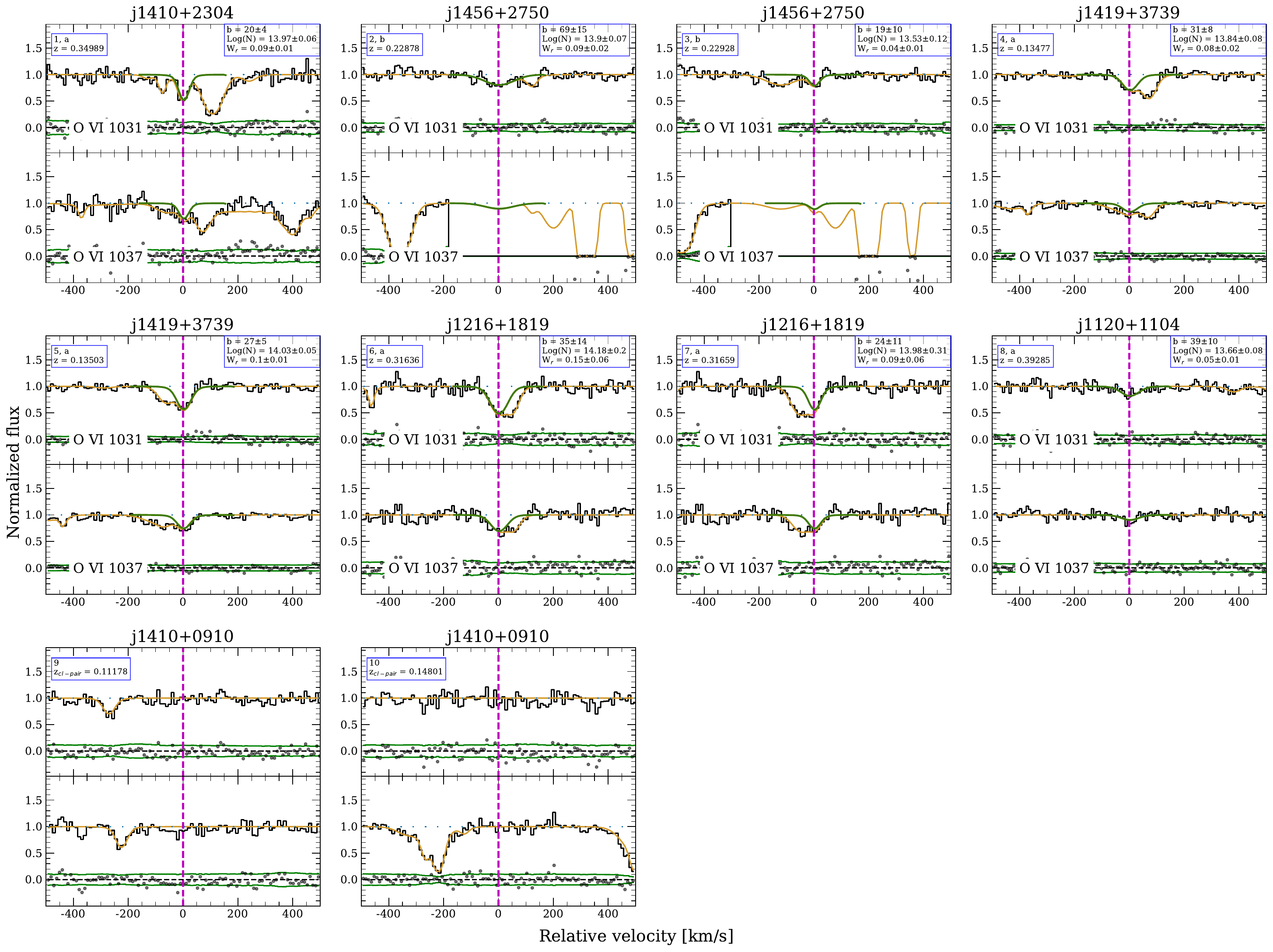}
  \caption{Oxygen transitions (\ion{O}{VI} $\lambda\lambda$ $1031.9$, $1037.6$ \AA) in relative velocity for each \ion{O}{VI} absorption line detected in our survey. We have included only absorption lines detected with reliability `a' or `b' and within a rest-frame velocity window of $\Delta v=\pm$ 1000 \kms and at impact parameter $\Delta d\leq$ 3 Mpc from the cluster-pairs. The spectrum is shown in black, the continuum in blue dots, and the Voigt profile fit in \ovicolor. The observational parameters, Doppler parameter in \kms, column density $\log$(N/cm$^{-2}$), and rest-frame equivalent width $W_r$ in \AA, are shown in the upper right part of the first plot in each panel. The detection ID, reliability, and redshift are shown in the upper-left corner of the first plot in each panel.}
  \label{Fig:Survey_OVI_transition}
\end{figure*}

In Figure~\ref{Fig:Survey_OVI_transition}, we show oxygen transitions (\ion{O}{VI} $\lambda\lambda$ $1031.9$, $1037.6$ \AA) for each \ion{O}{VI} absorption line detected in our survey with reliability `a' or `b'. We have included only absorption lines detected within a rest-frame velocity window of $\Delta v=\pm$ 1000 \kms and at impact parameter $\Delta d\leq$ 3 Mpc from the cluster-pairs. The spectrum is shown in black, the continuum in blue dots, and the Voigt profile fit in \ovicolor. The observational parameters, Doppler parameter in \kms, column density $\log$(N/cm$^{-2}$), and rest-frame equivalent width $W_r$ in \AA, are shown in the upper right part of the first plot in each panel. The detection ID, reliability, and redshift are shown in the upper-left corner of the first plot in each panel. Notice that for QSO J1456+2750, a gap in the spectrum is identified at the wavelength corresponding to the \ion{O}{VI} 1037.6 \AA\,transition.

\section{Redshift sensitivity function of \ion{H}{I} and \ion{O}{VI}.}
\label{sec:appendix_g_z}

\begin{figure*}
     \begin{subfigure}[b]{0.5\textwidth}
         \centering
         \includegraphics[width=\textwidth]{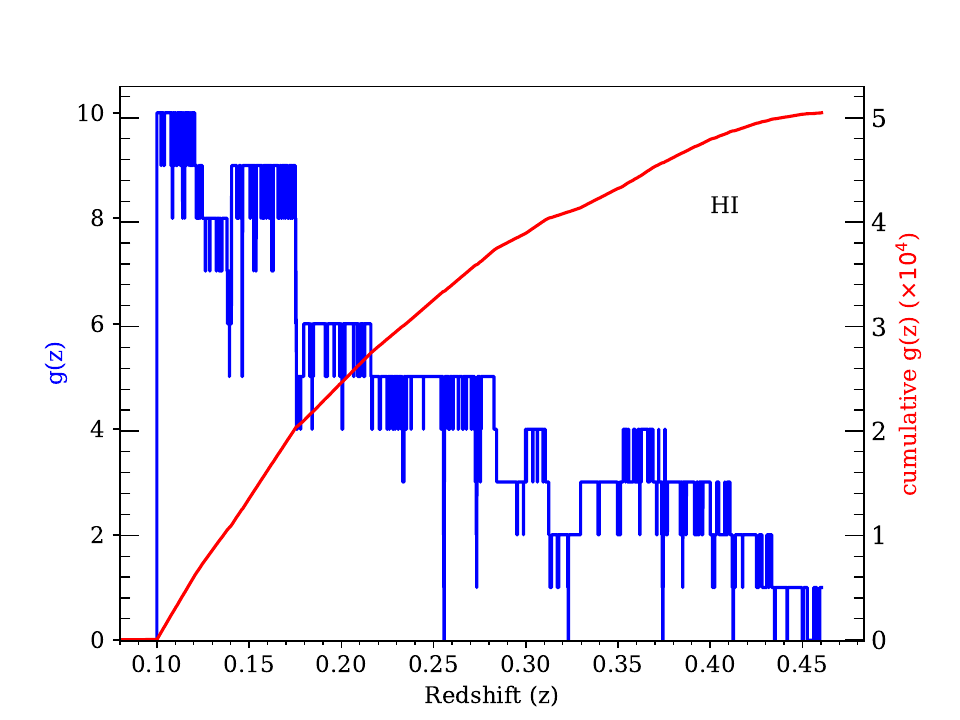}
     \end{subfigure}
     \hfill
     \begin{subfigure}[b]{0.5\textwidth}
         \includegraphics[width=\textwidth]{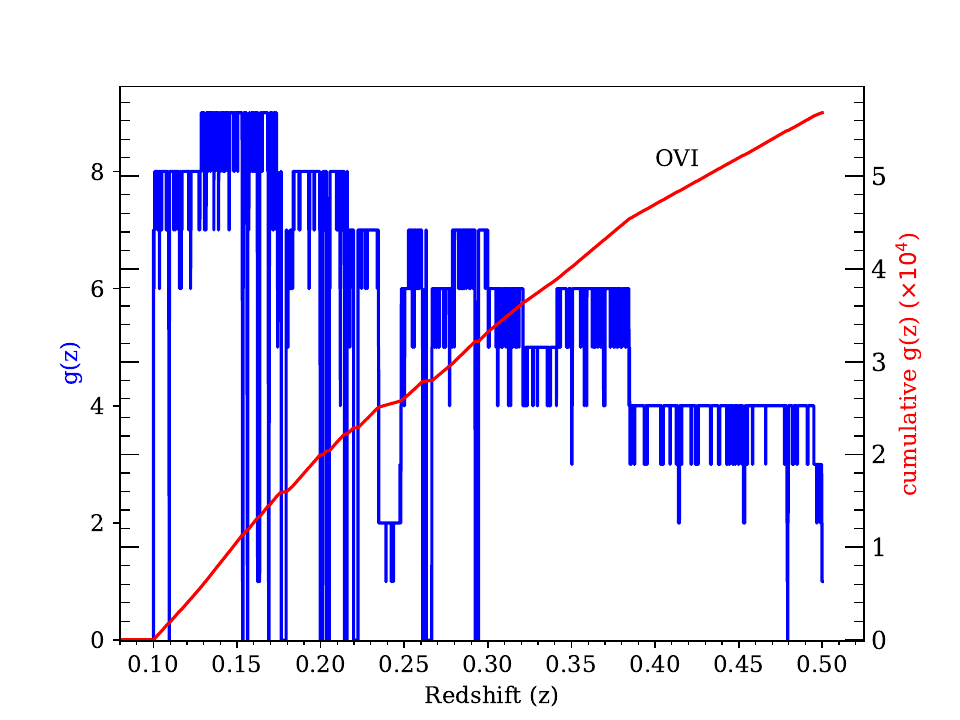}
     \end{subfigure}
      \caption{Full sightline redshift path sensitivity function versus redshift for $W_{\min}=0.03$ \AA. The sensitivity and cumulative functions are shown in blue and red, respectively. Left: \ion{H}{I}; right: \ion{O}{VI}.}
         \label{Fig:g_z_HI_OVI}
\end{figure*}

The left panel of Figure~\ref{Fig:g_z_HI_OVI} depicts the redshift path sensitivity function ($g(z)$) for \ion{H}{I} (in blue) and its cumulative curve (in red), considering the minimum equivalent width of $W_{min} = 0.03$ \AA. In the right panel, we show the redshift path sensitivity function for \ion{O}{VI}. The $g(z)$ function is represented in a histogram, and the bins have the size of a single pixel of the spectrum. This function indicates the coverage at low redshifts, in the range $0.1\leq z\leq 0.5$ for both ions, where our transitions are observed. For instance, Figure~\ref{Fig:g_z_HI_OVI} shows that in our survey, at redshifts $\lesssim 0.12$, we can find \ion{H}{I} in the ten QSO sightlines, but \ion{O}{VI} in only eight QSO sightlines. The $g(z)$ for \ion{H}{I} indicates that in our QSO survey, the maximum redshift at which \ion{H}{I} may be detected is $\sim0.46$, which is in agreement with the details provided in Section~\ref{2.3}. Additionally, we find that in our survey, there is a greater total coverage of \ion{O}{VI} than of \ion{H}{I}, as shown by the cumulative distribution plotted in red, where the maximum value for \ion{O}{VI} exceeds that for \ion{H}{I}.

\section{Representation of the position of each independent cluster-pair at $\Delta v = \pm500$ \kms}
\label{sec:appendix_geometry}

\begin{figure}
\centering
\includegraphics[width=0.49\textwidth]{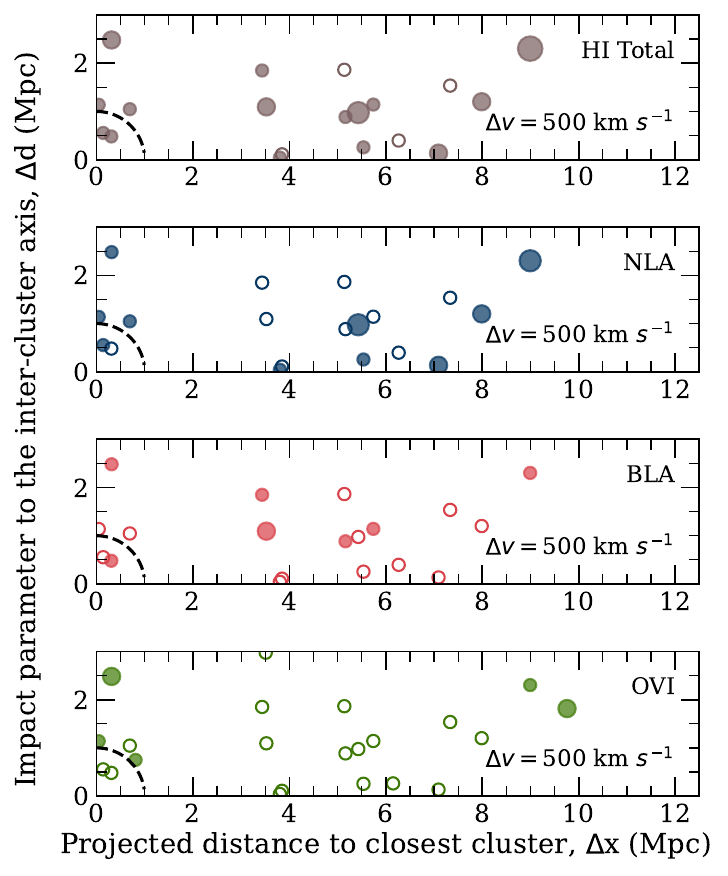}
  \caption{Same as Figure~\ref{Fig:deltax_deltad} but for a $\Delta v = \pm500$ \kms.}
  \label{Fig:geometry_appendix}
\end{figure}

Figure ~\ref{Fig:geometry_appendix} presents the location of the 22 independent cluster-pairs of our survey relative to their corresponding QSO sightline and their corresponding associated absorption features. An absorption line is associated with an independent cluster-pair if it is detected within $\Delta v = \pm500$ \kms, relative to its corresponding QSO sightline. The y-axis indicates the impact parameter to a given cluster-pair, $\Delta d$; while the x-axis indicates the distance of the closest galaxy cluster in a given cluster-pair to the QSO sightline, known as $\Delta x$. Additionally, the approximate virial radii expected for the sample of clusters are shown as a quarter dashed circumference of radii $1$ Mpc, centered at the origin. Filled symbols correspond to sightlines showing absorption line samples of the ion species indicated in each panel within $\Delta v$ from the redshifts of the corresponding cluster-pair. The size of the symbols is proportional to the number of detected absorption lines. Empty symbols correspond to sightlines showing no absorbers.

\section{Covering fraction}
\label{sec:appendix_covering}
\begin{figure}
\centering
\includegraphics[width=0.49\textwidth]{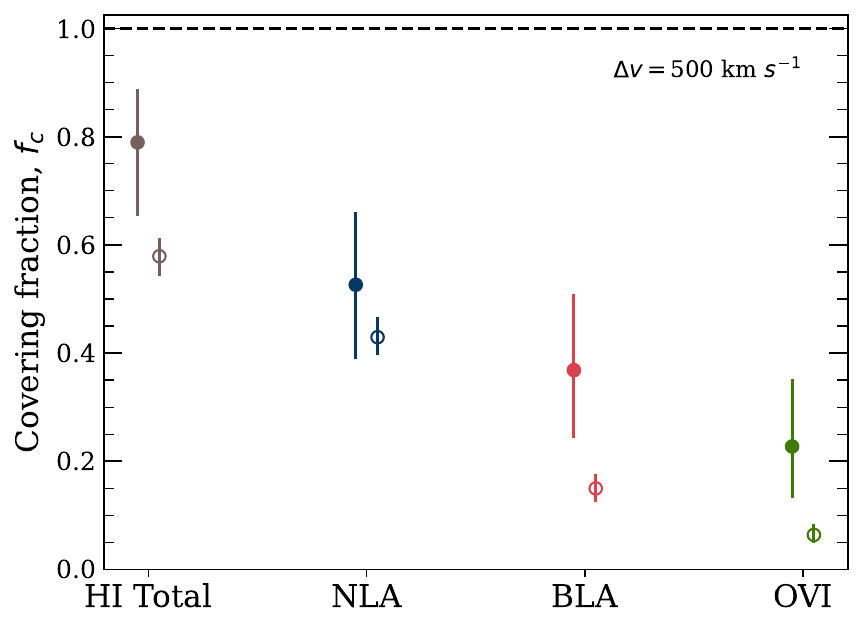}
  \caption{Same as Figure~\ref{Fig:covering_fraction} but for a $\Delta v = \pm500$ \kms.}
  \label{Fig:covering_appendix}
\end{figure}

Figure ~\ref{Fig:covering_appendix} shows the covering fraction for samples of total \ion{H}{I}, NLAs, BLAs, and \ion{O}{VI}, with the same colors as in the previous plots, within a rest-frame velocity window of $\Delta v=\pm500$ \kms around independent cluster-pairs. The $f_{c}^{filament}$ is shown in solid symbols and their corresponding randomly expected values, $f_{c}^{random}$, in open symbols. In Table~\ref{tab:appendix_covering_fraction_1000_500}, we compare the covering fractions found in our survey for each sample with the random expectation for $\Delta v=\pm1000$ \kms and $\Delta v=\pm500$ \kms.

\begin{table} 
\caption{Covering fractions of gas close to independent cluster-pairs for samples of total \ion{H}{I}, NLAs, BLAs, and \ion{O}{VI}, within a rest-frame velocity window of $\Delta v=\pm1000$ \kms (top) and $\Delta v=\pm500$ \kms (right) around independent cluster-pairs}
\label{cf}
 \centering 
 \renewcommand{\arraystretch}{1.5}
 \begin{tabular}{cccccc} 
 \hline\hline
Sample & $n_{hits}$ & $n_{trials}$ & $f_{c}^{filament}$ & $f_{c}^{random}$ & Excess \\
$(1)$ & $(2)$ & $(3)$ & $(4)$ & $(5)$ & $(6)$ \\ 
\hline
 \multicolumn{6}{c}{$\Delta v=\pm1000$ \kms and $\Delta d\leq3$ Mpc} \\
 \hline
Total \ion{H}{I}& $17$ & $19$ & $0.89_{-0.12}^{+0.07}$ & $1.00_{-0.01}^{+0.00}$ & $0.9$ \\ 
NLA& $14$ & $19$ & $0.74_{-0.14}^{+0.11}$ & $0.87_{-0.03}^{+0.02}$ & $0.9$ \\ 
BLA& $10$ & $19$ & $0.53_{-0.14}^{+0.13}$ & $0.30_{-0.03}^{+0.03}$ & $1.7$ \\ 
\ion{O}{VI}& $5$ & $22$ & $0.23_{-0.09}^{+0.12}$ & $0.13_{-0.02}^{+0.02}$ & $1.8$ \\  
\hline 
 \multicolumn{6}{c}{$\Delta v=\pm500$\kms and $\Delta d\leq3$ Mpc} \\
 \hline
Total \ion{H}{I}& $15$ & $19$ & $0.79_{-0.14}^{+0.10}$ & $0.58_{-0.04}^{+0.03}$ & $1.4$ \\ 
NLA& $10$ & $19$ & $0.53_{-0.14}^{+0.13}$ & $0.43_{-0.03}^{+0.04}$ & $1.2$ \\ 
BLA& $7$ & $19$ & $0.37_{-0.12}^{+0.14}$ & $0.15_{-0.03}^{+0.03}$ & $2.5$ \\ 
\ion{O}{VI}& $5$ & $22$ & $0.23_{-0.09}^{+0.12}$ & $0.06_{-0.02}^{+0.02}$ & $3.6$ \\  
\hline 
 \end{tabular} 
 \tablefoot{(1) Transition of absorbing gas. (2) Number of independent cluster-pairs showing absorption of a given transition. (3) Total number of independent-cluster-pairs where we can detect transitions. (4) Covering fraction close to cluster-pairs for each transition. (5) Covering fraction for a random sightline, estimated using Equation~\eqref{eq:fc_random}. (6) Excess covering fraction, calculated as $f_c^{filament}/f_c^{random}$.} 
 \label{tab:appendix_covering_fraction_1000_500}
 \end{table}

\end{appendix}
\end{document}